\documentclass[twocolumn,aps,pra,showpacs]{revtex4-1}

\usepackage{amssymb}
\usepackage{mathrsfs}
\usepackage{lipsum}
\usepackage{graphicx}
\usepackage[caption=false]{subfig}
\usepackage{mathtools}
\usepackage{epstopdf}
\usepackage{xcolor}
\usepackage{float}
\usepackage[colorlinks=true]{hyperref}
\DeclareGraphicsExtensions{.pdf,.jpg,.png,.eps}
\usepackage[toc,page,title,titletoc,header]{appendix}
\begin{document}

\title{Deterministic super-resolved estimation towards angular displacements based upon a Sagnac interferometer and parity measurement}

\author{Jian-Dong Zhang}
\affiliation{Department of Physics, Harbin Institute of Technology, Harbin, 150001, China}
\author{Zi-Jing Zhang}
\email[]{zhangzijing@hit.deu.cn}
\affiliation{Department of Physics, Harbin Institute of Technology, Harbin, 150001, China}
\author{Longz-Zhu Cen} 
\affiliation{Department of Physics, Harbin Institute of Technology, Harbin, 150001, China}
\author{Jun-Yan Hu} 
\affiliation{Department of Physics, Harbin Institute of Technology, Harbin, 150001, China}
\author{Yuan Zhao}
\email[]{zhaoyuan@hit.deu.cn}
\affiliation{Department of Physics, Harbin Institute of Technology, Harbin, 150001, China}

\date{\today}

\begin{abstract}
Super-resolved angular displacement estimation is of crucial significances for quantum information process and optical lithography.	
Here we report on and experimentally demonstrate a protocol for angular displacement estimation based on a coherent state containing orbital angular momentum. 
In the lossless scenario, with using parity measurement, this protocol can theoretically achieve 4$\ell$-fold super-resolution with quantum number $\ell$, and shot-noise-limited sensitivity saturating the quantum Cram\'er-Rao bound. 
Several realistic factors and their effects are considered, including nonideal state preparation, photon loss, and imperfect detector.
Finally, given mean photon number $\bar N=2.297$ and $\ell=1$, we show an angular displacement super-resolution effect with a factor of 7.88, and the sensitivity approaching shot-noise limit is reachable.

\end{abstract}

\pacs{03.65.Wj, 42.50.Ar, 06.20.-f, 03.67.Bg}

\maketitle

\section{introduction}

As is known to all, a light beam can carry two forms of angular momenta: spin angular momentum (SAM), and orbital angular momentum (OAM).
SAM corresponds to the polarization of the light, and the angular momentum of each photon is $\sigma \hbar $, where $\sigma  = {\rm{ + }}1$ and $\sigma  =  - 1$ stand for the left-handed and the right-handed polarized light, respectively. 
OAM is associated with the azimuthal distribution of the light, and each photon in the OAM beam carries an angular momentum $\ell\hbar $ with quantum number $\ell$. 
The polarization of the light is discovered early and is applied among a great deal of fields, feature detections, target imaging, and material identifications, to name a few \cite{song2014polarization, israel2014supersensitive, kartazayeva2005backscattering}. 
The fact that a beam with an azimuthal phase dependence of $\exp \left( {i\ell\theta } \right)$ is capable of carrying OAM is put forward by Allen $et$ $al$. in 1992 \cite{allen1992orbital}. 
Since then, plenty of theoretical and experimental studies have focused on this subject \cite{simpson1997mechanical, tabosa1999optical, padgett2002orbital}. 
The infinite orthogonal dimensions of OAM place no limit on the amount of information that can be carried by a single photon. 
Therefore, within past decade, OAM has played a significant role in the fields of quantum communications, quantum computing, and quantum metrology \cite{yan2014high, nicolas2014quantum, puentes2012weak}. 
Due to the characteristic of helical phase $\exp \left( {i\ell\theta } \right)$,  an OAM state can take the part of `angular amplifier', which converts an angular displacement $\theta$ into the amplified displacement $\ell\theta$ \cite{d2013photonic}.

Phase shifts and optical rotations, the fundamental operations for photonic qubit gates, are two vital degrees of freedom in parameter estimation. 
For the past few years, phase estimation has been discussed in numerous physical protocols \cite{taylor2016quantum, giovannetti2011advances}, many exotic quantum states and measurement strategies are presented.
Angular displacement estimation is also widely analyzed in quantum process tomography \cite{zhou2015quantum} and weak measurement \cite{magana2014amplification, thekkadath2016direct, zhang2015precision}, however, it is seldom mentioned in interferometric quantum metrology.
On the other hand, almost all quantum states are sensitive to photon loss, and they are also limited by difficult preparation for large photon number.
In the scenario of high loss channel, these facts downplay the advantages arising from quantum resources, and coherent states come across as ideal candidate.   
In this paper, we demonstrate a novel estimation protocol towards angular displacements using a coherent state and a Sagnac interferometer (SI) combined with a Dove prism.

The remainder of this paper is organized as follows. 
In Sec. \ref{s2}, we introduce the fundamental principle and measurement strategy of our protocol.
Section \ref{s3} focuses on studying the effects of several realistic factors on our protocol, such as nonideal state preparation, photon loss, and imperfect detector.
In Sec. \ref{s4}, we discuss the fundamental sensitivity limit of our protocol by calculating quantum Fisher information (QFI), and compare it with the previous Mach-Zehnder interferometer (MZI) protocol.
An experimental realization is demonstrated in Sec. \ref{s5}, and the performance is briefly analyzed. 
Finally, we summarize our work in Sec. \ref{s6}.

\section{Fundamental principle and measurement strategy of the protocol }
\label{s2}

\begin{figure}[!ht]
\centering
\includegraphics[width=8cm]{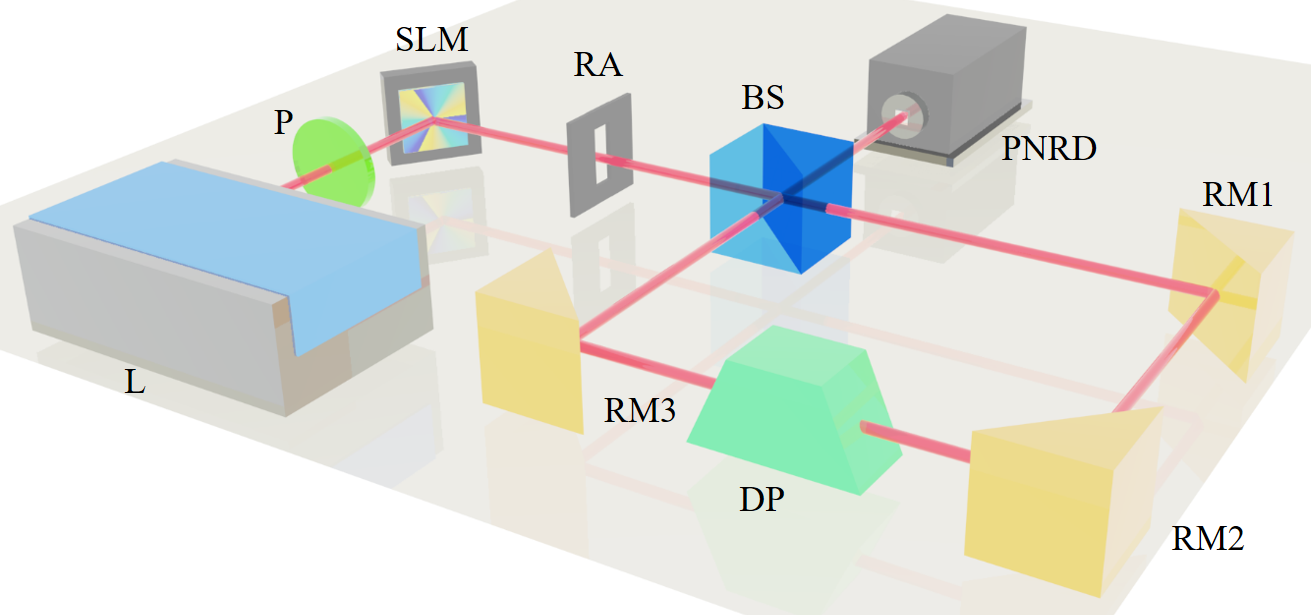}
\caption{Schematic of the angular displacement estimation protocol. The full names of the abbreviations in the figure: L, laser; P, polarizer; SLM, spatial light modulator; RA, rectangular aperture; BS, beam splitter; DP, Dove prism; RM, reflection mirror; PNRD, photon-number-resolving detector.}
\label{f1}
\end{figure}

To begin, let us consider the angular displacement estimation protocol of which the setup is a SI consisting of three mirrors and a 50/50 beam splitter arranged in a square, as illustrated in Fig. \ref{f1}. 
The coherent state is generated from a laser, and its OAM degree of freedom is added by a spatial light modulator \cite{fickler2012quantum, zhou2016orbital}. 
The polarizer is used to filter the polarization which is not suitable for the spatial light modulator, and rectangular aperture is responsible for passing the first-order diffraction.
In accordance with the theory of quantum optics, the input state can be described as $\left| {{\psi _\textrm{in}}} \right\rangle  = {\left| {{\alpha _\ell}} \right\rangle _A}{\left| 0 \right\rangle _B}$, where ${\alpha _\ell}{\rm{ = }}\sqrt N $ and $N$ is the mean photon number in the coherent state. 
Then, the input enters the SI and is divided into two paths, in turn, the state becomes $\left| {{\psi _1}} \right\rangle  = {\left| {{{{\alpha _\ell}} \mathord{\left/
				{\vphantom {{{\alpha _\ell}} {\sqrt 2 }}} \right.
				\kern-\nulldelimiterspace} {\sqrt 2 }}} \right\rangle _A}{\left| {{{i{\alpha _\ell}} \mathord{\left/
				{\vphantom {{i{\alpha _\ell}} {\sqrt 2 }}} \right.
				\kern-\nulldelimiterspace} {\sqrt 2 }}} \right\rangle _B}$.
Here we assume that the clockwise direction in the interferometer loop is path $A$, and the counterclockwise one is path $B$. 
The beams in two paths pass through the Dove prism \cite{leach2002measuring} with an angular displacement $\varphi$, which is the parameter we would like to estimate.
After such an evolution process, the state can be expressed as $\left| {{\psi _2}} \right\rangle  = {\left| {{{{\alpha _\ell}{e^{i2\ell\varphi }}} \mathord{\left/
 {\vphantom {{{\alpha _\ell}{e^{i2\ell\varphi }}} {\sqrt 2 }}} \right.
 \kern-\nulldelimiterspace} {\sqrt 2 }}} \right\rangle _A}{\left| {{{i{\alpha _\ell}{e^{ - i2\ell\varphi }}} \mathord{\left/
 {\vphantom {{i{\alpha _\ell}{e^{ - i2\ell\varphi }}} {\sqrt 2 }}} \right.
 \kern-\nulldelimiterspace} {\sqrt 2 }}} \right\rangle _B}$. 
Finally, the state goes through the beam splitter again, and the output has the following ket representation $\left| {{\psi _\textrm{out}}} \right\rangle  = {\left| {i{\alpha _\ell}\cos \left( {2\ell\varphi } \right)} \right\rangle _A}{\left| { - i{\alpha _\ell}\sin \left( {2\ell\varphi } \right)} \right\rangle _B}$.

Comparing with the MZI, our protocol has two main advantages. 
On the one hand, the beams in two paths are more likely to experience identical optical paths and photon losses in our protocol, in that SI is a self-balanced interferometer.  
On the other hand, our protocol is equivalent to the scenario that the two Dove prisms in the two paths of MZI are reversely rotated with same angular displacement.

We turn now to the measurement strategy.
Parity measurement is originally discussed by Bollinger $et$ $al.$ for enhanced frequency measurement with an entangled state\cite{bollinger1996optimal}, subsequently, Gerry and Campos apply it to optical interferometers \cite{gerry2000heisenberg, gerry2005quantum}. 
Generally, the implementation of parity measurement  requires a photon-number resolving detector, and the details of the detector can be found in Refs. \cite{achilles2004photon, cohen2014super, liu2017fisher}. 
In this strategy, the counts are assigned as $ + 1$ and $ - 1$ for even and odd photon numbers, respectively. 
Therefore, the parity operator for output port $B$ can be written as $ {\hat \Pi }  = \exp \left( {i\pi {{\hat b}^\dag }\hat b} \right)$.

In the Fock basis, the output state is recast as
\begin{eqnarray}
\nonumber\left| {{\psi _\textrm{out}}} \right\rangle  = &&{e^{ - \frac{1}{2}{N}}}\sum\limits_{n = 0}^\infty  {\frac{{{{\left[ {i{\alpha _\ell}\cos \left( {2\ell\varphi } \right)} \right]}^n}}}{{\sqrt {n!} }}}  \\ 
&&\times\sum\limits_{m = 0}^\infty\frac{{{{\left[ { - i{\alpha _\ell}\sin \left( {2\ell\varphi } \right)} \right]}^m}}}{{\sqrt {m!} }}\left| {n,m} \right\rangle. 
\label{1}
\end{eqnarray}
Further, the probability of simultaneously detecting $n$ photons at port $A$ and $m$ ones at port $B$ is
\begin{equation}
P\left( {n,m} \right) = \frac{{{e^{ - {N}}}}}{{n!m!}}{\left[ {{N}{{\cos }^2}\left( {2\ell\varphi } \right)} \right]^n}{\left[ {{N}{{\sin }^2}\left( {2\ell\varphi } \right)} \right]^m}.
\label{2}
\end{equation}
Consequently, the conditional probability ${P_{\rm even}}$ or ${P_{\rm odd}}$ for port $B$ can be calculated through a series sum of $P\left( {n,m} \right)$ over the parity of the photon number $n$,
\begin{eqnarray}
{P_\textrm{even}} &&= \frac{1}{2}\left\{ {1 + \exp \left[ { - 2{N}{{\sin }^2}\left( {2\ell\varphi } \right)} \right]} \right\}, 
\label{3}\\ 
{P_\textrm{odd}} &&= \frac{1}{2}\left\{ {1 - \exp \left[ { - 2{N}{{\sin }^2}\left( {2\ell\varphi } \right)} \right]} \right\}. 
\label{4}
\end{eqnarray}

In the light of the definition of the parity operator, we can obtain the expectation value of the output,
\begin{equation}
\left\langle {\hat \Pi } \right\rangle  = \exp \left[ { - 2N{{\sin }^2}\left( {2\ell\varphi } \right)} \right].
\label{5}
\end{equation}
Further, with the help of error propagation, the sensitivity is given by
\begin{equation}
\Delta \varphi  = \frac{{\sqrt {\left\langle {{{\hat \Pi }^2}} \right\rangle  - {{\left\langle {\hat \Pi } \right\rangle }^2}} }}{{\left| {{{\partial \left\langle {\hat \Pi } \right\rangle } \mathord{\left/
					{\vphantom {{\partial \left\langle {\hat \Pi } \right\rangle } {\partial \varphi }}} \right.
					\kern-\nulldelimiterspace} {\partial \varphi }}} \right|}}
= \frac{{\sqrt {\exp \left[ {4{N}{{\sin }^2}\left( {2\ell\varphi } \right)} \right] - 1} }}{{\left| {4\ell{N}\sin \left( {4\ell\varphi } \right)} \right|}}.
\label{6}
\end{equation}
By using first-order approximation, when $\varphi$ approaches 0, the sensitivity arrives at its minimum,
\begin{equation}
\Delta {\varphi _{\min }} = {\left. {\frac{{\sqrt {1 + 4N{{\sin }^2}\left( {2\ell\varphi } \right) - 1} }}{{\left| {4\ell N\sin \left( {4\ell\varphi } \right)} \right|}}} \right|_{\varphi  \to 0}} = \frac{1}{{4\ell\sqrt N }}.
\label{7}
\end{equation}

\begin{figure}[!ht]
\centering
\includegraphics[width=7cm]{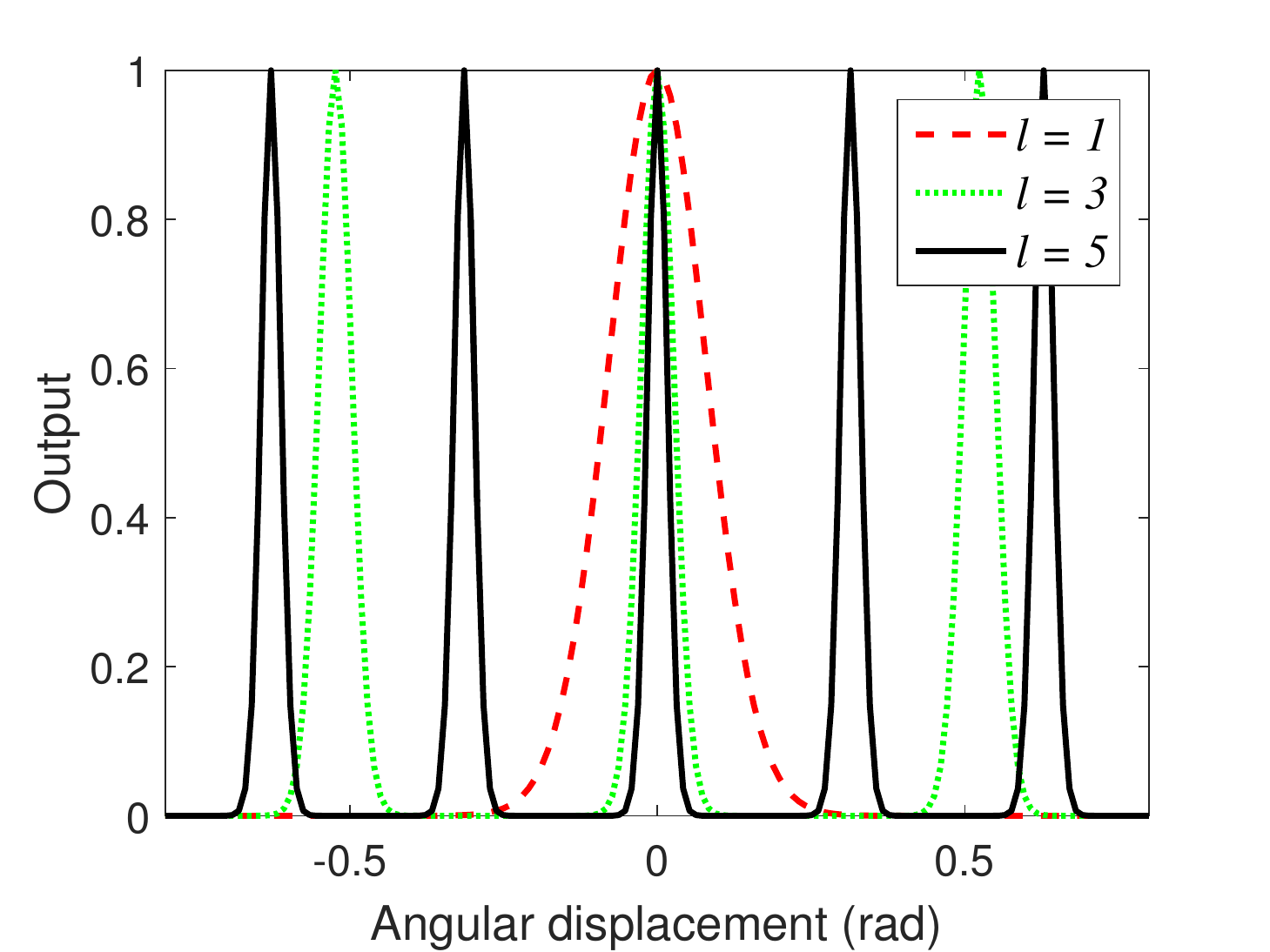}
\includegraphics[width=7cm]{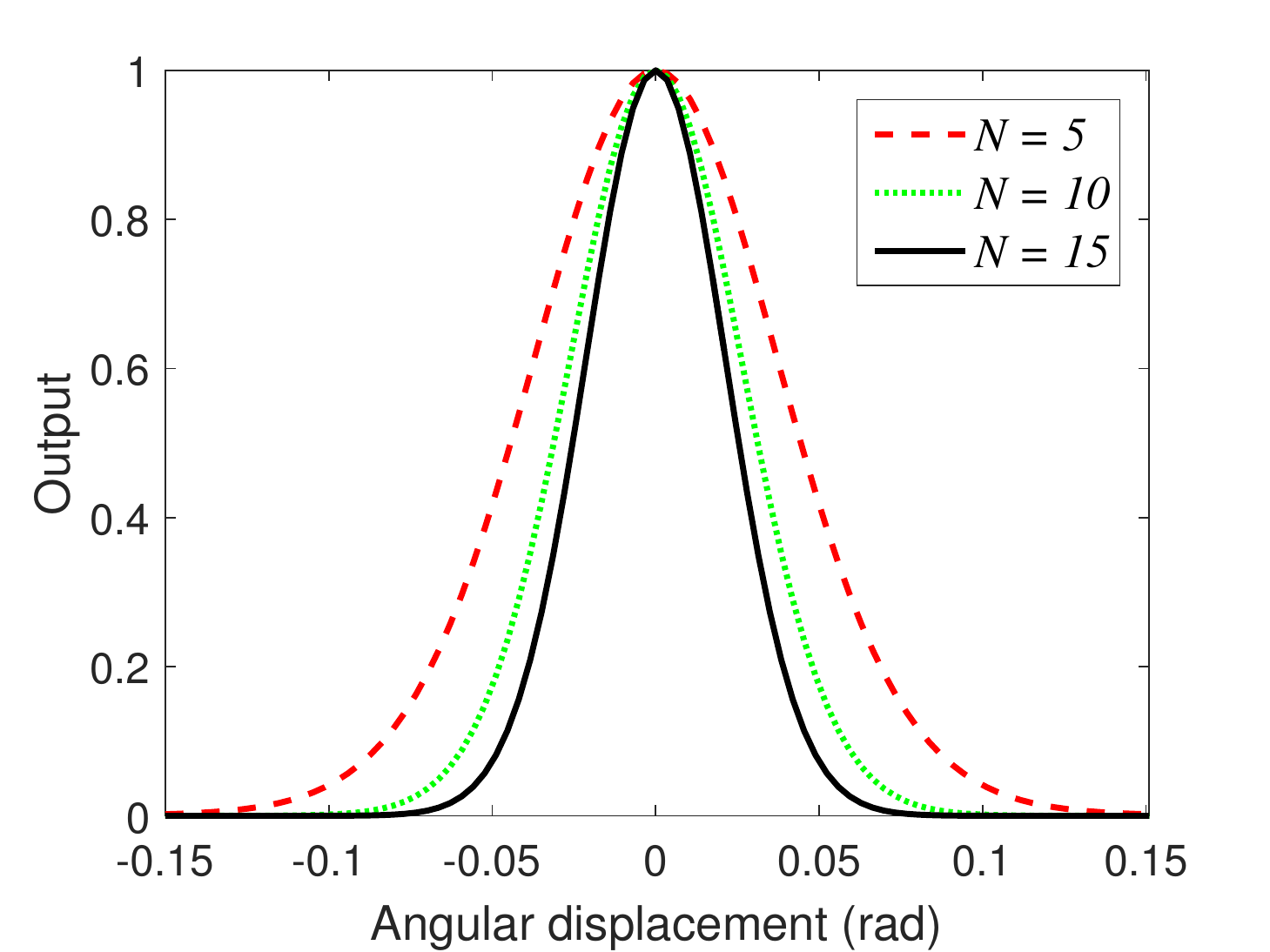}
\caption{The resolution of parity measurement as a function of angular displacement. (a) $N =10$. (b) $\ell =3$.}
\label{f2}
\end{figure}

In the above derivation, we have used a property of the parity operator, $ {{{\hat \Pi }^2}}  = 1$. 
For intuitively observing the variation on the resolution caused by mean photon number $N$ and quantum number $\ell$, we plot Fig. \ref{f2}. 
From the figure we can find that the number of oscillating output fringes increase with increasing $\ell$, and each fringe gets narrow as the increase of $N$.  
Hence, the resolution of the protocol can be improved with an increasing value of either $N$ or $\ell$. 
Moreover, the visibility of the output is approximate to 100\%, in that the maximum sits at 1 and the minimum approaches 0 for large $N$.
The definition of visibility refers to \cite{N00N}
\begin{equation}
V = \frac{{{{\left\langle {\hat \Pi } \right\rangle }_{\max }} - {{\left\langle {\hat \Pi } \right\rangle }_{\min }}}}{{{{\left\langle {\hat \Pi } \right\rangle }_{\max }} + {{\left\langle {\hat \Pi } \right\rangle }_{\min }}}}.
\label{8}
\end{equation}

\begin{figure}[!ht]
\centering
\includegraphics[width=7cm]{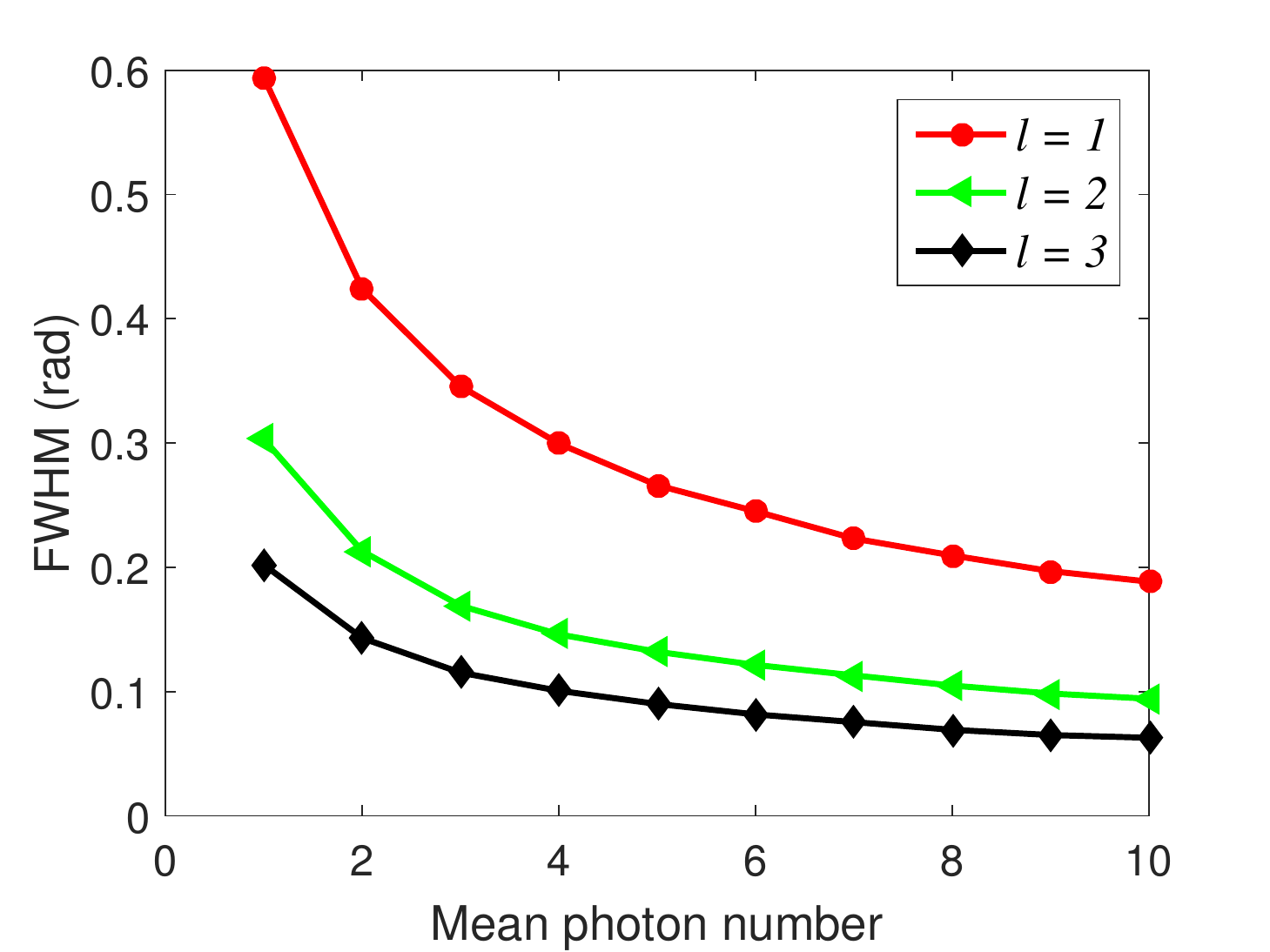}
\caption{The FWHM of parity measurement as a function of mean photon number.}
\label{f3}
\end{figure}

In Fig. \ref{f3}, we show the full widths at half maximum (FWHMs) with different values of $N$ and  $\ell$. 
FWHM is a universal super-resolution criterion, i.e., the smaller the FWHM is, the higher the resolution is. 
Figure \ref{f3} indicates that the increase of $N$ or $\ell$ can provide an enhancement of the resolution, and a more apparent resolution increase is obtained whenever both $N$ and $l$ are increased.  
With respect to the sensitivity, Eq. (\ref{7}) shows a shot-noise-limited sensitivity as the factor $4\ell$ is a classical effect.
The effect arising in OAM is equal to the increase of the number of trials.
The results mean that the increases of both $N$ and $\ell$ have an enhanced effect on the sensitivity. 

\section{Effects of realistic factors}
\label{s3}

Since a setup is inevitably immersed in its surrounding environment, the realistic factors will affect the estimation.
In this section, we analyze the effects of several realistic factors on the resolution and the sensitivity.
These factors may take place in three stages: state preparation; state evolution; and state measurement.
Each stage will be discussed in this section, and the subscript $k$ stands for the $k$-th realistic factor.

\subsection{Nonideal state preparation}

We start off with the nonideal state preparation.
In this scenario, the input must be described by a density matrix, rather than a state vector \cite{bahder2011phase, zhang17effects}. 
Let us assume that the conversion efficiency of the spatial light modulator is $\eta$, consequently, we can write the input density matrix as
\begin{equation}
{\rho _\textrm{in}} = \left[ {\eta \left| {{\alpha _\ell}} \right\rangle \left\langle {{\alpha _\ell}} \right| + \left( {1 - \eta } \right)\left| {{\alpha _0}} \right\rangle \left\langle {{\alpha _0}} \right|} \right] \otimes \left| 0 \right\rangle \left\langle 0 \right|.
\label{9}
\end{equation}

In the light of the evolution process mentioned, the reduced output density matrix for mode $B$ can be obtained.
Further, the expectation value of parity operator is
\begin{equation}
{\left\langle {\hat \Pi } \right\rangle _{1}} = {\rm Tr}\left( {\hat \Pi {\rho _{B\textrm{out}}}} \right) = \eta\exp \left[ { - 2 {N}{{\sin }^2}\left( {2\ell\varphi } \right)} \right]+1-\eta,
\label{10}
\end{equation}
using first-order approximation again, we have the optimal sensitivity,
\begin{equation}
\Delta {\varphi _2} = {\left. {\frac{{\sin \left( {2l\varphi } \right)\sqrt {1 - \eta N{{\sin }^2}\left( {2l\varphi } \right)} }}{{\left| {2l\sqrt {\eta N} \sin \left( {4l\varphi } \right)} \right|}}} \right|_{\varphi  \to 0}} = \frac{1}{{\sqrt \eta  }}\frac{1}{{4l\sqrt N }}
\label{11}
\end{equation}
Equation (\ref{9}) suggests that only the photons added with OAM degree of freedom play a role in measurement, and the unmodulated photons boost the minimum of the output. 
This nonideal efficiency reduces the visibility since the minimum is boosted, in turn, the sensitivity is also deteriorating by a factor of ${\sqrt \eta }$.

\subsection{Photon loss}

We next take into account the effect of a type of inevitable realistic factor, photon loss, on the resolution and the sensitivity. 
The measured information in the output is acquired through counting the results of multiple trials. 
The lossy photon in each measurement is random, however, the statistical results are subject to a certain probability distribution. 
In general, the theoretical simulation of photon loss is realized by inserting a virtual beam splitter in the interference loop, the transmissivities of the two paths are $\sqrt {{T_A}} $ and $\sqrt {{T_B}} $ \cite{feng2014quantum, kacprowicz2010experimental}. 
Further, the parameters $L_A=1-T_A$ and $L_B=1-T_B$ represent two path losses.
On the basis of this theory, the output state for mode $B$ reduces to $\left| {{\psi _\textrm{out}}} \right\rangle_B  = {\left| {{{{\alpha _\ell}\left( {\sqrt {{T_B}} {e^{ - i2\ell\varphi }} - \sqrt {{T_A}} {e^{i2\ell\varphi }}} \right)}}/{2}} \right\rangle}$.
The expression of the resolution and the sensitivity corresponding to this output state can be calculated through the analysis,
\begin{equation}
{\left\langle {\hat \Pi } \right\rangle _{2}}  = {\kern 1pt}  \exp \left[ {N\sqrt {{T_A}{T_B}} \cos \left( {4\ell\varphi } \right)}{ - \frac{N}{2}\left( {{T_A} + {T_B}} \right)} \right] 
\label{12}
\end{equation}
and
\begin{eqnarray}
\nonumber{\Delta \varphi _{2}}  = && \sqrt {\exp \left\{ {N\left[ {{T_A} + {T_B} - 2\sqrt {{T_A}{T_B}} \cos \left( {4\ell\varphi } \right)} \right]} \right\} - 1}  \\ 
&&\times \frac{1}{{\left| {4\ell\left( {{T_A} + {T_B}} \right)N\sin \left( {4\ell\varphi } \right)} \right|}}.  
\label{13}
\end{eqnarray}

\begin{figure}[htbp]
\centering
\includegraphics[width=7cm]{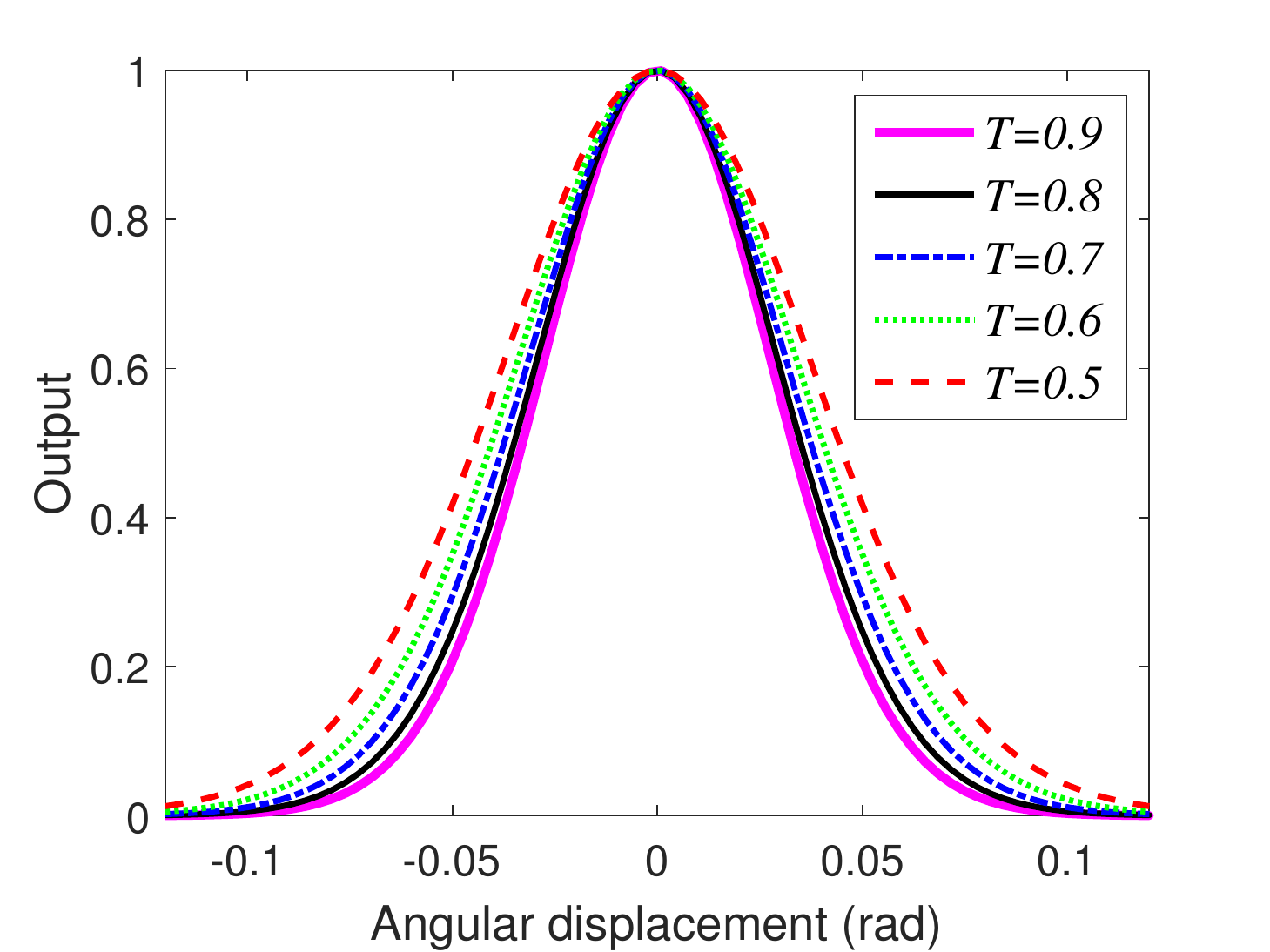}
\includegraphics[width=7cm]{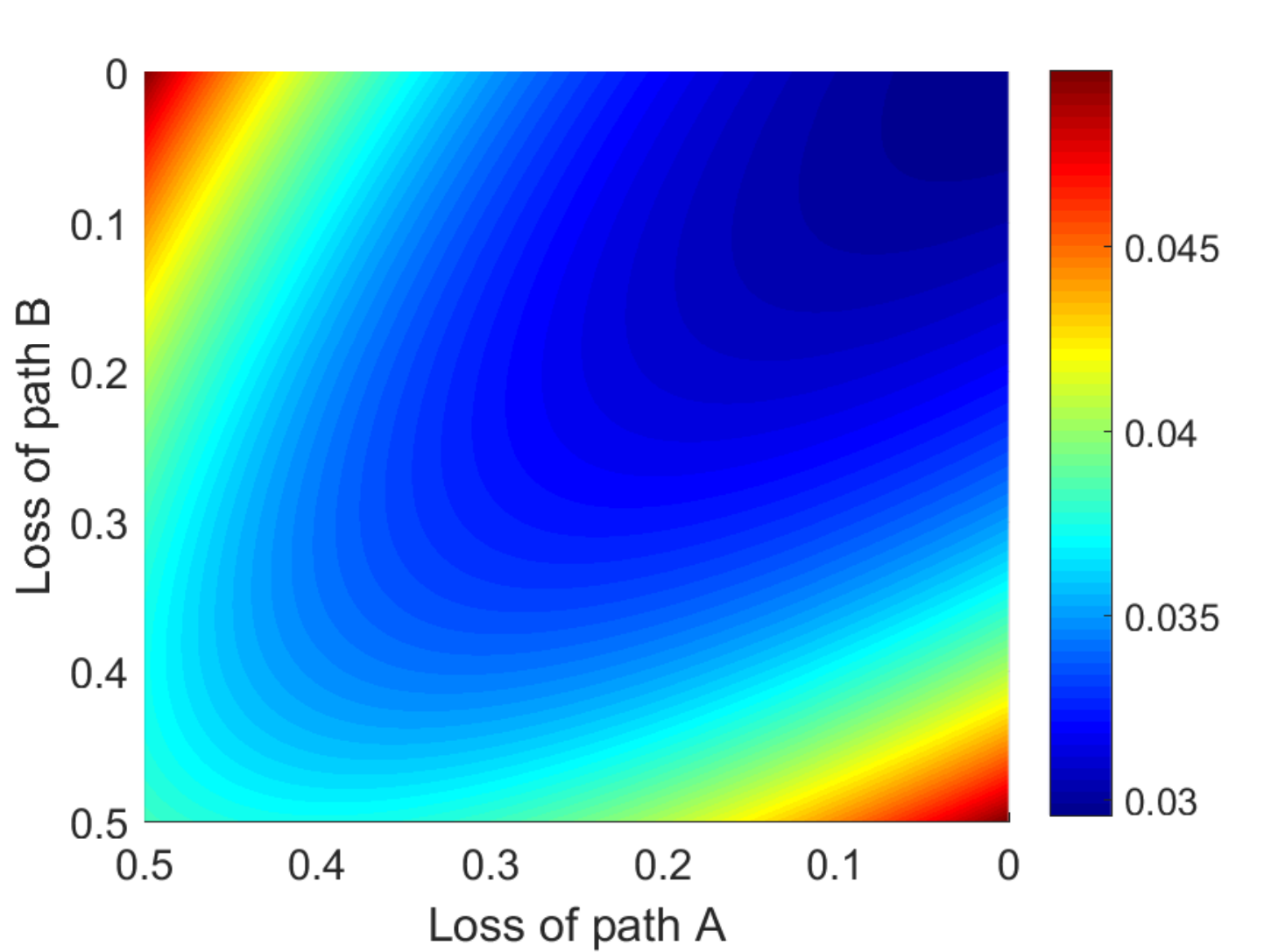}
\caption{(a) The resolution of parity measurement as a function of angular displacement, where $N=10$. (b) The sensitivity of parity measurement as a function of two path losses, where $N=10$ and $\ell=3$.}
\label{f4}
\end{figure}

From Eq. (\ref{12}) we can find that only if the condition ${T_A} = {T_B}$ is satisfied does the maximum of output take on the value of 1. 
This is consistent with the interference condition of classical optics. 
Meanwhile, for the SI protocol, ${T_A} \approx {T_B}$ is facile to be satisfied as the two light fields experience the same path. 
Figure \ref{f4}(a) manifests the FWHM is broadening as the two transmissivities decrease, and the visibility remains changeless. 
As for the sensitivity, Fig. \ref{f4}(b) indicates that, for the identical total loss, the optimal sensitivity is achieved under the same photon losses in two paths.

\subsection{Imperfect detector}

Detection efficiency, response-time delay and dark counts are three typical imperfect factors of the detector \cite{spagnolo2012phase}.
Specific results of the analysis are as follows.

\subsubsection{Detection efficiency}

In general, there is no guarantee that a detector keeps a 100\% efficiency, and this process is also simulated by inserting a virtual beam splitter in front of the detector, where the transmissivity is $\kappa$ \cite{PhysRevA.83.063836}, also known as detection efficiency.
The output state for mode $B$ can be rewritten as ${\left| {{\psi _\textrm{out}}} \right\rangle _B} = \left| { - i\sqrt \kappa  {\alpha _l}\sin \left( {2\ell\varphi } \right)} \right\rangle $, and the expectation value is
\begin{equation}
{\left\langle {\hat \Pi } \right\rangle _{3}} = \exp \left[ { - 2\kappa {N}{{\sin }^2}\left( {2\ell\varphi } \right)} \right].
\label{14}
\end{equation}

One can find that this equation is the same as Eq. (\ref{12}) when $T_A=T_B=\kappa$. 
This shows that the effect of the detection efficiency is identical with that of the photon loss, and the previous conclusions are still applicable.
This phenomenon stems from the fact that photon loss is a linear loss, that is, a coherent state maintains its distribution under linear loss, the presence of loss after the SI completely equals a lossless SI fed by a weaker input \cite{PhysRevLett.99.223602}.

\subsubsection{Response-time delay and dark counts}

In the practical measurements, response-time delay and dark counts also affect the performance of the detector.
The former forces the width of the sampling detection gate to increase, as a result, the rate of the latter will rise, and a detailed analysis of this process is available in Appendix \ref{A}.
A thoughtful discussion of the effect of the dark counts on the output with parity measurement has been proposed in Ref. \cite{huang2017adaptive}.
Here we invoke this conclusion, and the expectation value of the parity operator equals
\begin{equation}
{\left\langle {\hat \Pi } \right\rangle _{4}} = {e^{ - 2r}}\left\langle {\hat \Pi } \right\rangle, 
\label{15}
\end{equation}
where the parameter $r$ is the rate of the dark counts. 
With the help of error propagation, the sensitivity can be calculated.   
Under the current technology, the range of $r$ is generally between ${10^{ - 8}}$ to ${10^{ - 3}}$.

\begin{figure}[H]
\centering
\includegraphics[width=7cm]{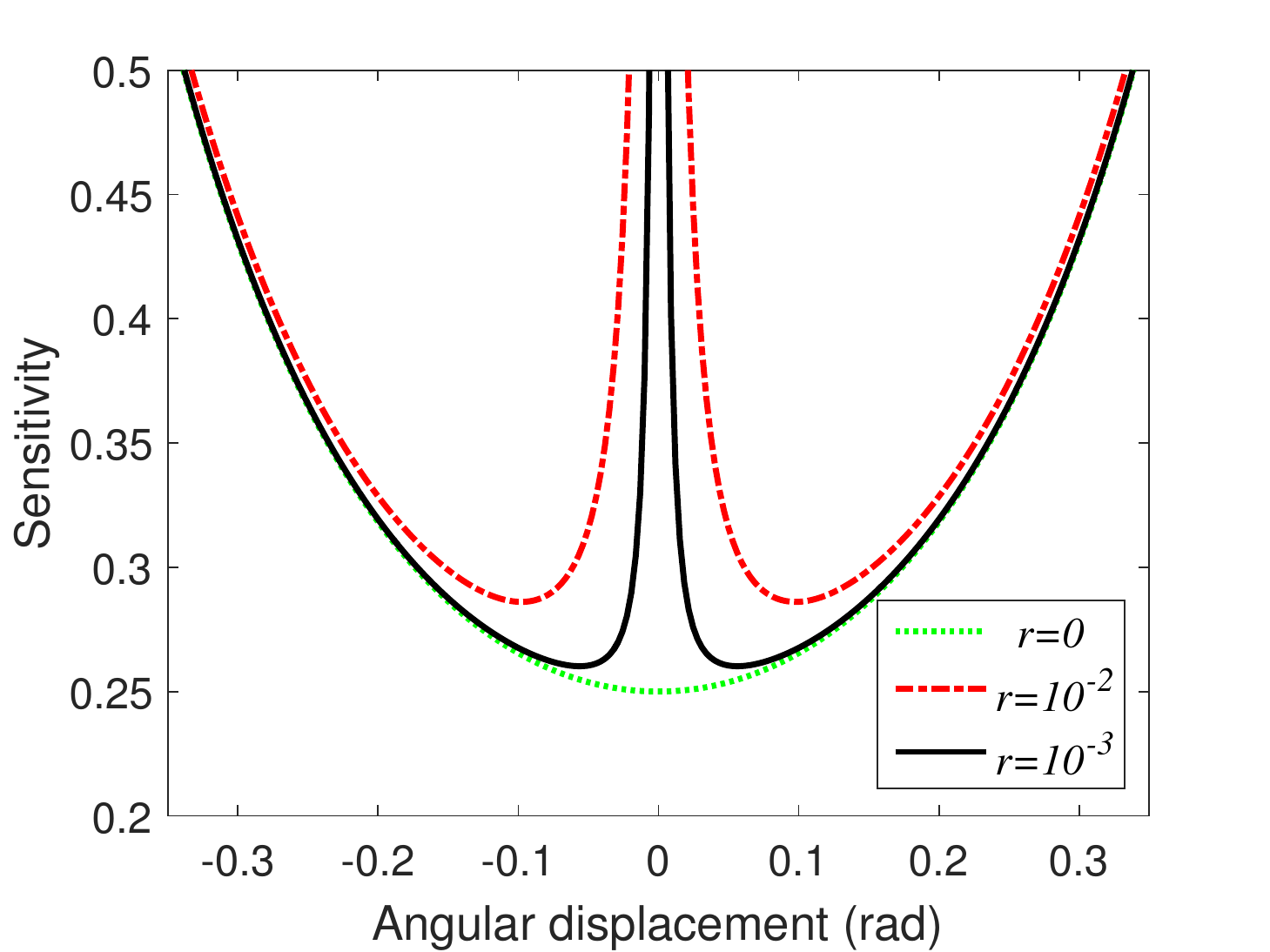}
\caption{The effects of the response-time delay and the dark counts on the sensitivity. Where the curve of $r=0$ is the ideal curve, the curves of $r=10^{-3}$ and $r=10^{-2}$ respectively correspond to the scenarios: only dark counts; and the combination of dark counts and the response-time delay.}
\label{f8}
\end{figure}

The impact of response-time delay on dark counts will increase the rate of dark counts by a factor which is generally less than 10. 
Hence, we choose ${10^{ - 3}}$ and ${10^{ - 2}}$ to represent the scenarios: only dark counts; and the combination of dark counts and the response-time delay.
The results in Fig. \ref{f8} exhibit that the change of sensitivity is slight with only dark counts, however, the deterioration of sensitivity becomes obvious due to the simultaneous existence of the response-time delay and dark counts.

\section{Analysis of fundamental sensitivity limit}
\label{s4}

In the above sections, we merely calculate the sensitivity of measurement strategy, the fundamental sensitivity limit over all possible positive operate valued measures (POVMs) is not given. 
Here we systematically compare the our protocol and previous MZI protocol from the perspective of QFI.

\begin{figure}[htbp]
\centering
\includegraphics[width=8cm]{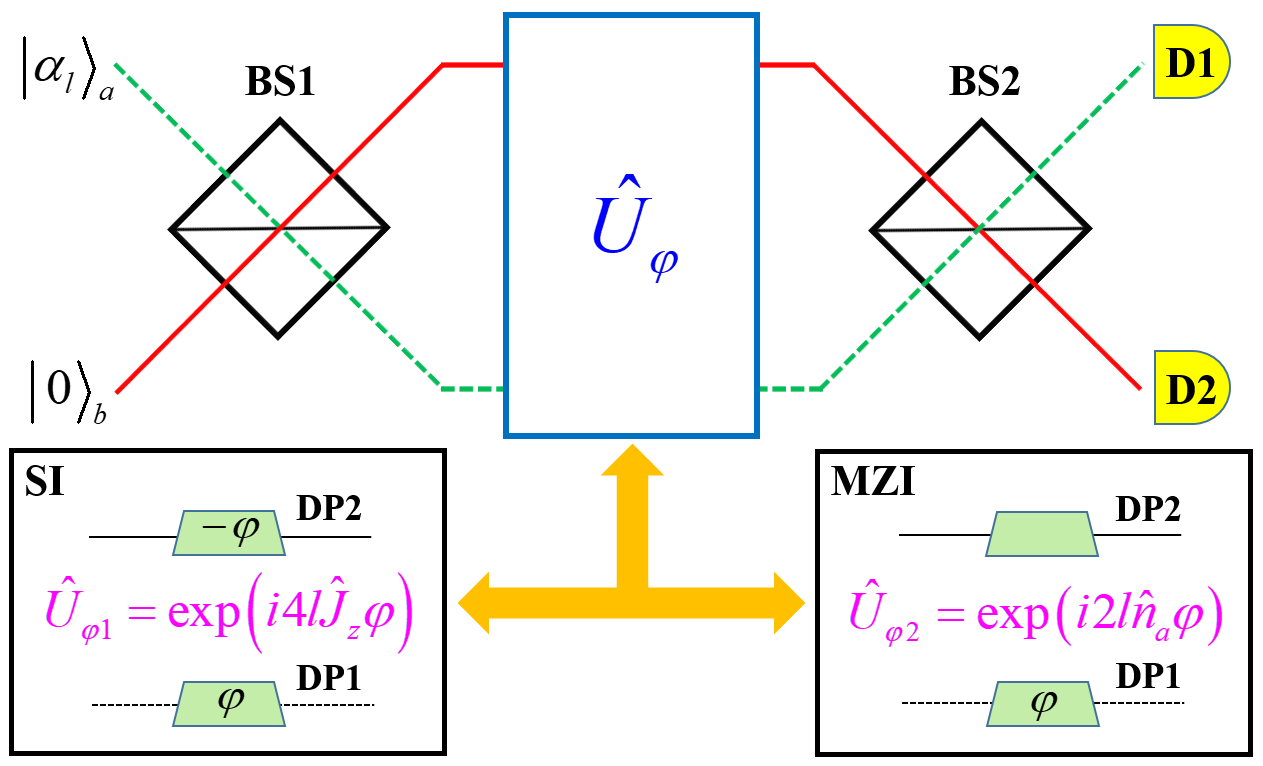}
\caption{Diagram of the angular displacement estimation protocol. The full names of the abbreviations in the figure: DP, Dove prism; D, detector; BS, beam splitter; SI, Sagnac interferometer; MZI, Mach-Zehnder interferometer.}
\label{f6}
\end{figure}

The current angular displacement estimation protocols can be divided into the following two categories: SI and MZI protocols, as illustrated in the Fig. \ref{f6}. 
For the above two protocols, the rotation of Dove prism can be described as the following operators ${\hat U_{\varphi 1}} = \exp \left( {i4\ell{{\hat J}_z}\varphi } \right)$ and ${\hat U_{\varphi 2}} = \exp \left( {i2\ell{{\hat n}_a}\varphi } \right)$, respectively. 
The operator for beam splitter is ${\hat U_\textrm{BS}} = \exp \left( {i{\pi }{{\hat J}_x}/2} \right)$, where
\begin{eqnarray}
 {{\hat J}_x} &&= \frac{1}{2}\left( {{a^\dag }b + a{b^\dag }} \right), \\ 
 {{\hat J}_y} &&=  - \frac{i}{2}\left( {{a^\dag }b - a{b^\dag }} \right), \\ 
 {{\hat J}_z} &&= \frac{1}{2}\left( {{a^\dag }a - {b^\dag }b} \right) 
\label{17}
\end{eqnarray}
are the angular momentum operators in the Schwinger representation \cite{tan2014enhanced}. 
These operators satisfy the cyclic commutation relations for the Lie algebra of SU(2): $\left[ {{{\hat J}_x} {\kern 1pt} {\kern 1pt} ,{\kern 1pt}  {{\hat J}_y}} \right] = i{\hat J_z}$; $\left[ {{{\hat J}_y}{\kern 1pt} {\kern 1pt} , {\kern 1pt} {\kern 1pt} {{\hat J}_z}} \right] = i{\hat J_x}$; and $\left[ {{{\hat J}_z}{\kern 1pt} {\kern 1pt} ,{\kern 1pt} {\kern 1pt} {{\hat J}_x}} \right] = i{\hat J_y}$. 
The input density matrix can be written as ${\rho _\textrm{in}} = {\rho _a} \otimes {\rho _b}$, where ${\rho _a} = \left| {{\alpha _\ell}} \right\rangle \left\langle {{\alpha _\ell}} \right|$ and ${\rho _b} = \left| 0 \right\rangle \left\langle 0 \right|$. 
Here we define the counterclockwise path in Fig. \ref{f6} is mode $a$, and the clockwise one is mode $b$.

In accordance with the above analysis, we calculate the QFI for two scenarios. 
As for SI protocol, the output density matrix evolves into ${\rho _\textrm{out}} = {\hat U_{\varphi 1}}{\hat U_\textrm{BS}}{\rho _\textrm{in}}\hat U_\textrm{BS}^\dag \hat U_{\varphi 1}^\dag $, and in terms of the equation ${{\partial {\rho _\textrm{out}}} \mathord{\left/
 {\vphantom {{\partial {\rho _\textrm{out}}} {\partial \varphi }}} \right.
 \kern-\nulldelimiterspace} {\partial \varphi }} = {{ - i\left[ {{\rho _\textrm{out}},{\kern 1pt} \hat R} \right]} \mathord{\left/
 {\vphantom {{ - i\left[ {{\rho _\textrm{out}},{\kern 1pt} {\kern 1pt} {\kern 1pt} \hat R} \right]} \hbar }} \right.
 \kern-\nulldelimiterspace} \hbar }$, we can obtain the estimator $\hat R$. 
For the case of a pure state input, the QFI is simplified as $4{\left\langle {{\Delta ^2}\hat R} \right\rangle _\textrm{in}}$ \cite{gibilisco2007uncertainty, knysh2011scaling, liu2013phase}. 
Therefore, the QFI for SI is calculated as
\begin{eqnarray}
\nonumber{{\cal F}_\textrm{SI}} && = 4\left[ {\left\langle \psi  \right|{{\left( {4\ell{{\hat J}_z}} \right)}^2}\left| \psi  \right\rangle  - \left\langle \psi  \right|4\ell{{\hat J}_z}{{\left| \psi  \right\rangle }^2}} \right] \\ 
&&= 16{\ell^2}{N}.
\label{19}
\end{eqnarray}

In the above derivation, we have used the formula ${\hat U_\textrm{BS}}\left| \alpha  \right\rangle \left| 0 \right\rangle  = \left| {{\alpha  \mathord{\left/
 {\vphantom {\alpha  {\sqrt 2 }}} \right.
 \kern-\nulldelimiterspace} {\sqrt 2 }}} \right\rangle \left| {{{i\alpha } \mathord{\left/
 {\vphantom {{i\alpha } {\sqrt 2 }}} \right.
 \kern-\nulldelimiterspace} {\sqrt 2 }}} \right\rangle  \equiv \left| \psi  \right\rangle $. 
And the relationship between the optimal sensitivity $\Delta \varphi_{\rm min} $ and QFI is $\Delta \varphi_{\rm min}  = {1 \mathord{\left/
 {\vphantom {1 {\sqrt {qF} }}} \right.
 \kern-\nulldelimiterspace} {\sqrt {\nu{\cal F}} }}$, where $\nu$ is the number of trials. 
The parameter $\nu$ does not lead any quantum effects into the sensitivity since it is a classical experiment repeat. 
In the next discussion we focus on single trial, that is, $\nu = 1$. 
The optimal sensitivity of the SI protocol can be calculated, $\Delta {\varphi _\textrm{SI}} = {1 \mathord{\left/
 {\vphantom {1 {4\ell\left| {{\alpha _l}} \right|}}} \right.
 \kern-\nulldelimiterspace} {4\ell\left| {{\alpha _\ell}} \right|}}$.

For the scenario of the MZI protocol, we have
\begin{eqnarray}
\nonumber{{\cal F}_\textrm{MZI}} && = 4\left[ {\left\langle \psi  \right|{{\left( {2\ell{{\hat a}^\dag }\hat a} \right)}^2}\left| \psi  \right\rangle  - \left\langle \psi  \right|2\ell{{\hat a}^\dag }\hat a{{\left| \psi  \right\rangle }^2}} \right] \\
&&= 8{\ell^2}{N}.
\label{20}
\end{eqnarray}

It is obvious that the QFI of the SI protocol is superior to that of the MZI one, i.e., SI protocol is more sensitive to angular displacement.

An interesting and perplexing phenomenon with respect to ${{\cal F}_{\rm MZI}}$ is that the optimal sensitivity corresponding to Eq. (\ref{20}) is $\Delta {\varphi _\textrm{MZI}} = {1 \mathord{\left/
 {\vphantom {1 {2\sqrt 2 \ell\left| {{\alpha _l}} \right|}}} \right.
 \kern-\nulldelimiterspace} {2\sqrt 2 \ell\left| {{\alpha _\ell}} \right|}}$. 
After removing the factor 2$\ell$ originating from OAM, the QFI also implies a sub-shot-noise-limited sensitivity. 
In order to solve this confusion, we can use the phase-averaging approach to ascertain whether a measurement strategy can break through the shot-noise limit with only a coherent state in the absence of additional source. 
This approach can give a low-down on sensitivity limit with only using input source. 
A simplified understanding for its idea is to disrupt the input state into a mixed state losing all phase references. 
Based on this approach, we obtain the QFI of the MZI protocol, ${{\cal F}_{\bar \rho }} = \sum\nolimits_{{{n = 0}}}^\infty  {{p_n}4{\ell^2}} n = 4{\ell^2}N$.
This QFI implies the sensitivity limit is shot-noise limit, and the details can be found in Appendix \ref{B}. 
Hence, the QFI in Eq. (\ref{20}) contains a part of information stemming from additional sources.
Ascertaining the special additional sources which can assist MZI protocol to achieve the sensitivity in Eq. (\ref{20}) is still a meaningful and challenging research content, for many practical measurements can be classified as the MZI configuration.
Overall, the SI protocol is more sensitive for angular displacement estimation, and its sensitivity is twice as much as that of MZI protocol.

\section{Experimental realization}
\label{s5}

As the last part of the work in this paper, we perform the proof of principle with $\ell=1$. The working principle and measuring results about the photon-number-resolving detector are supplied in Appendix \ref{C}. 
As can be seen from Fig. \ref{f7}, the experimental results are in agreement with the theoretical analysis.
In Fig. \ref{7}(a) we fit the expectation value of the output in terms of experimental data, 
\begin{equation}
\left\langle {\hat \Pi } \right\rangle  = 0.9507\exp \left\{ { - 4.594{{\sin }^2}\left[ {2\left( {\varphi  - 0.7022} \right)} \right]} \right\}.
\label{21}
\end{equation}

\begin{figure}[htbp]
	\centering
	\includegraphics[width=7cm]{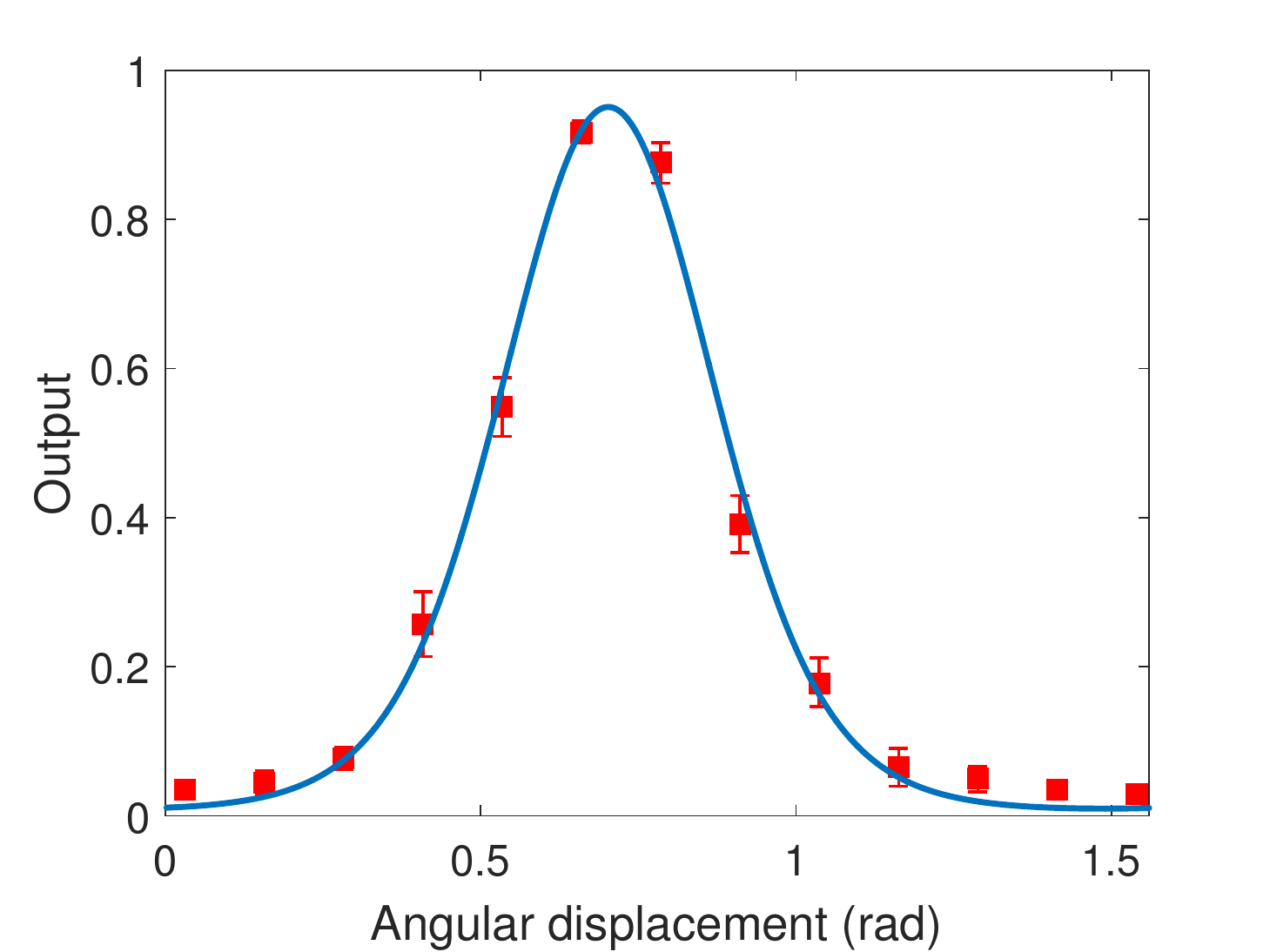}
	\includegraphics[width=7cm]{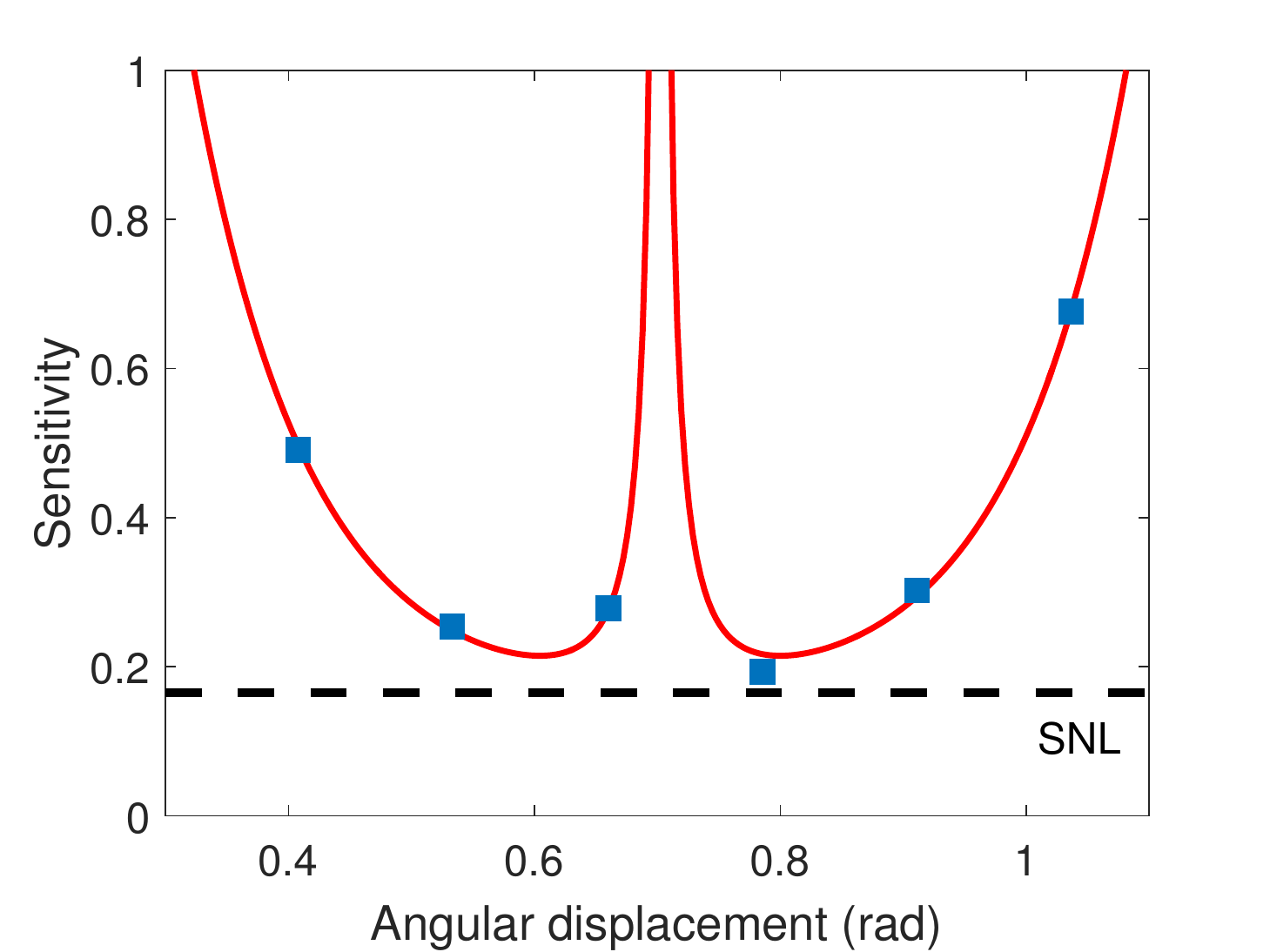}
	\caption{Experimental data as a function of angular displacement with $\ell=1$. (a) The blue line is a fit to the output. Error bars are one standard deviation due to propagated Poissonian statistics. (b) The red line is sensitivity deduced from the fit of the output, blue dots are the sensitivities calculated from the experimental data, and the black dashed line is the shot-noise limit defined in accordance with $\bar N$}
	\label{f7}
\end{figure}

This equation implies that the mean photon number arriving at the detector is $\bar N=2.297$, and the visibility of the output is 98\%.
Note that here $\bar N=T \kappa N$, the effect of photon loss is not reflected in Eq. (\ref{21}) as we only record the mean photon number arriving at the detector.

By calculating the FWHM, the experimental data demonstrates that our protocol has an enhanced resolution with a factor of 7.88. 
This also points out that our protocol can be applied to the field of optical lithography \cite{PhysRevLett.85.2733}.
Moreover, ignoring the relative position of the maximum, the Eq. (\ref{21}) can be recast as
\begin{equation}
\left\langle {\hat \Pi } \right\rangle  = \exp \left[ { - 4.594{{\sin }^2}\left( {2\varphi } \right) }-0.0506 \right].
\label{22}
\end{equation}
That is, the rate value $r$ is 0.0253 in the experiment, and comprises dark counts, response-time delay, and background noise.
These noises result that the maximum in Fig. \ref{f7}(a) cannot reach 1.

Figure \ref{7}(b) presents the sensitivities calculated from experimental data.
The results mean that the sensitivities tally with the theoretical analysis and the output fit.
Note that the optimal sensitivity is slightly inferior to the shot-noise limit due to the noise photons, and this scenario is similar to the discussion about the dark counts of imperfect detector.

\section{Conclusion}
\label{s6}

In conclusion, we introduce a novel interferometric setup, a SI with a Dove prism, which can realize super-resolved angular displacement estimation using parity measurement. 
The input state is a coherent state carrying OAM, and in lossless scenario we can obtain 4$\ell$-fold resolution fringe and shot-noise-limited sensitivity. 
The resolution and the sensitivity can be improved by increasing mean photon number and quantum number, independently or simultaneously. 
We also discuss the effects of several realistic factors on the performances of the output.
Nonideal preparation efficiency brings the deteriorations on the resolution, the visibility and the sensitivity.
With respect to photon loss, for identical total loss, the scenario that two same path losses provides a better resolution and an optimal sensitivity. 
The effects of dark counts and response-time delay on the sensitivity are unconspicuous, and the resolution is not affected by them.
Additionally, the fundamental sensitivity limits of our protocol and MZI one are given by calculating QFI, the results suggest that the sensitivity of SI protocol is saturated by QCR bound and is twice as much as that of MZI protocol.
Finally, a proof of principle is performed, the experimental data tally with the theoretical analysis.
For mean photon number $\bar N=2.297$, we achieve a super-resolved output which is enhanced by a factor of 7.88 and a nearly shot-noise-limited sensitivity.

\section*{Acknowledgments}
We would like to thank Prof. Zhi-Yuan Zhou and Shi-Long Liu from University of Science and Technology of China for a great deal of enlightening discussions with the experiment. 
This work is supported by the National Natural Science Foundation of China (Grant No. 61701139).

\appendix

\section{The effect of the response-time delay on the rate of dark counts}
\label{A}

Here we offer the explanation about the relationship between the response-time delay and dark counts.
In the practical measurements, the response-time delay can be expressed as a mean time delay attached to a delay jitter $\tau$. 
The mean delay has no effect on the estimation results, for the measurement strategy is to count the photon number, rather than arriving time. 
 
\begin{figure}[!ht]
\centering
\includegraphics[width=7.5cm]{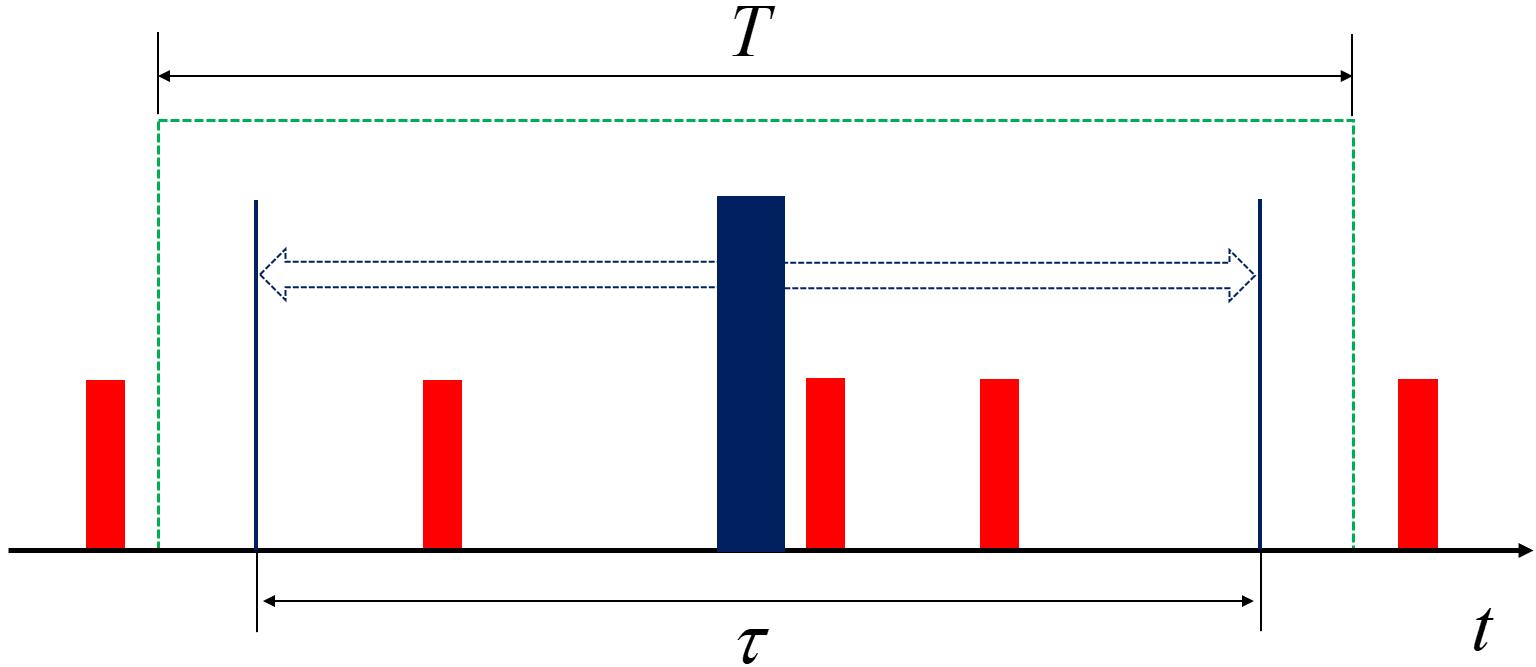}
\caption{Schematic of the effect of the response-time delay on the measurement results.}
\label{fs1}
\end{figure}

Schematic diagram for the effect of the response-time delay on the detection results is shown in Fig. \ref{fs1}, where the blue rectangle is the theoretical standard response-time. 
However, the time of the practical arriving signal may occur at any point in the $\tau$ of a period of time in the presence of the response-time delay. 
The parameter $T$ is the time width of sampling detection gate and the relationship $\tau \le T$ is satisfied to guarantee that only one signal in each gate. 
The red rectangle expresses the pulse of the dark counts of which the distribution is random and the statistical results follow the Poissonian distribution. 
Moreover, the dark counts outside the sampling detection gate do not affect the measurement. 
The width of the gate has to be increased owing to the response-time delay, hence, the effect of time delay on the measurement results is to increase the rate of dark counts.

\section{QFI of MZI protocol using phase-averaging approach}
\label{B}

In this part of Appendix, we give an elaborate calculation process for the phase-averaging method. 
In this framework, phase randomization is required for the input density matrix, i.e.,
\begin{eqnarray}
\nonumber{{\bar \rho }_1} =&& \frac{1}{{2\pi }}\int_0^{2\pi } {\exp \left( {i\delta {{\hat n}_a}} \right)} \exp \left( {i\delta {{\hat n}_b}} \right){\rho _a} \\
\nonumber&&\otimes {\rho _b}\exp \left( { - i\delta {{\hat n}_a}} \right)\exp \left( { - i\delta {{\hat n}_b}} \right)d\delta  \\ 
=&& \sum\limits_{n = 0}^\infty  {{p_n}\left| {{n}} \right\rangle \left\langle {{n}} \right| \otimes \left| 0 \right\rangle \left\langle 0 \right|}.
\label{B1}
\end{eqnarray}
Where ${p_n} = {{{{N}^{n}}\exp \left( { - {N}} \right)} / n}!$ is the probability of emerging $n$ photons in OAM coherent state $\left| {{\alpha _\ell}} \right\rangle $. 
It is easy to find that the off-diagonal elements of the density matrix disappear at this point, that is, the coherence information is erased.

Then the density matrix passes through the first beam splitter and becomes
\begin{eqnarray}
\nonumber{{\bar \rho }_2} &&= {{\hat U}_\textrm{BS}}{{\bar \rho }_1}\hat U_\textrm{BS}^\dag \\ 
&&= \sum\limits_{n = 0}^\infty  {p_n}\sum\limits_{m = 0}^n {C_n^m }  \left| {{n} - {m}} \right\rangle \left\langle {{n} - {m}} \right| \otimes \left| {{m}} \right\rangle \left\langle {{m}} \right| ,
\label{B2}
\end{eqnarray}
where $C_n^m$ is binomial coefficient. 
In view of the orthogonality of the Fock state ($\left\langle {n}
 {\left | {\vphantom {n m}}
 \right. \kern-\nulldelimiterspace}{m} \right\rangle  = {\delta _{nm}}$) and the convexity of the QFI, the QFI of the entire mixed state equals the sum of that of each Fock state in the light of the weight factor ${p_n}$.
For a two-mode Fock state $\left| {{n}} \right\rangle \left| 0 \right\rangle $ and a unitary evolution process ${\hat U_{\varphi 2}}{\hat U_{\rm BS}}$, its QFI can be calculated as
\begin{eqnarray}
\nonumber{{\cal F}_\textrm{Fock}} &&= 4\left[ {\left\langle {{{\hat U}_\textrm{BS}}{{\left( {2\ell{{\hat a}^\dag }\hat a} \right)}^2}\hat U_\textrm{BS}^\dag } \right\rangle  - {{\left\langle {{{\hat U}_\textrm{BS}}\left( {2\ell{{\hat a}^\dag }\hat a} \right)\hat U_\textrm{BS}^\dag } \right\rangle }^2}} \right]\\ 
 &&= 4{\ell^2}n.
\label{B3}
\end{eqnarray}
The expectation values are taken over the Fock state $\left| {{n},0} \right\rangle $, here we have used the Baker-Hausdorff lemma ${e^{ - i{\pi }{{\hat J}_x}/2}}{\hat{a}^\dag }a{e^{i{\pi }{{\hat J}_x}/2}} = {\hat J_y} + {\hat J_z}$ and the unitary property of the operator ${\hat U_\textrm{BS}}$. 
Consequently, the QFI of input density matrix in Eq. (\ref{B1}) goes to 
\begin{equation}
{{\cal F}_{\bar \rho }} = \sum\limits_{n = 0}^\infty  {{p_n}4{\ell^2}} n = 4{\ell^2}N.
\label{B4}
\end{equation}

Note that the above result is the shot-noise limit, that is, the optimal sensitivity of the MZI protocol is the shot-noise limit in the scenario of a coherent state input and without additional driving sources.

\section{The working principle and measuring results of the photon-number-resolving detector}
\label{C}

\begin{figure*}[htpb]
	\centering
	\includegraphics[width=7cm]{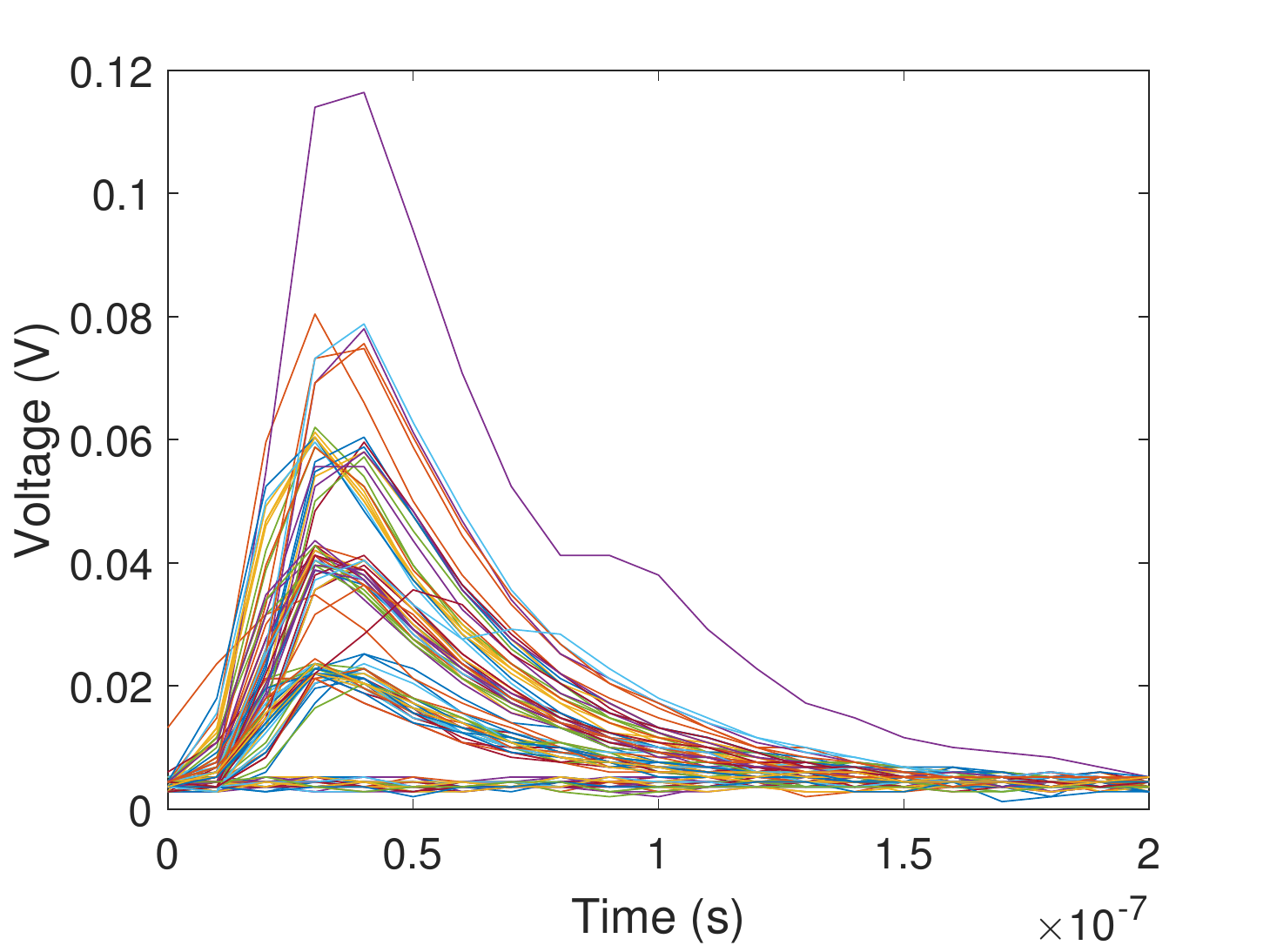}
	\includegraphics[width=7cm]{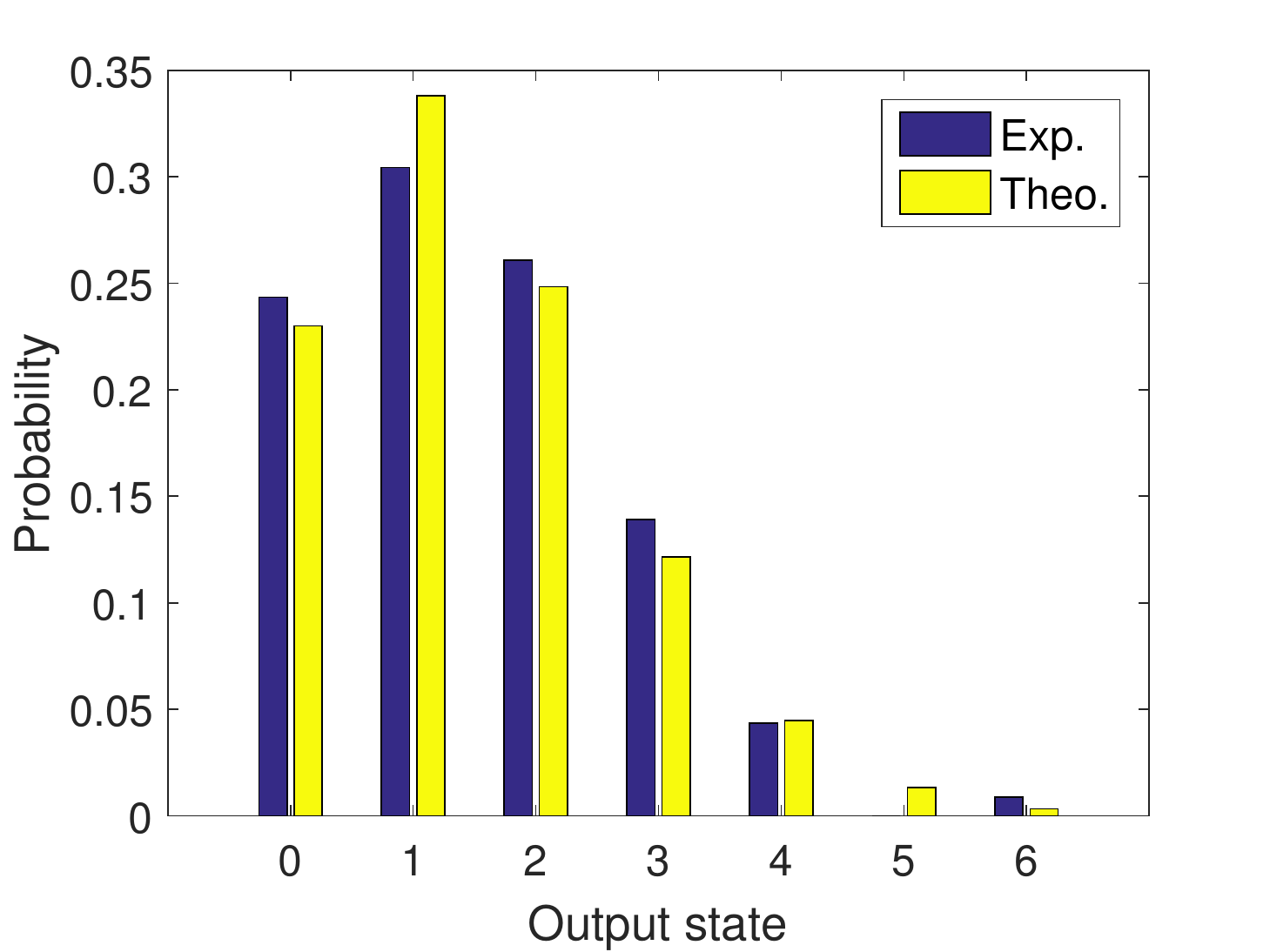}
	\caption{(a) The analog voltage signals displayed by oscillograph, and each signal is converted from single statistical trigger counts. (b) The probability distribution of output photon state and Poissonian distribution fit.}
	\label{f_detector}
\end{figure*}

The photon-number-resolving detector we used in experiment is a Geiger mode avalanche photodiode (Gm-APD) array.
Each APD only responses to the presence or absence of photons at output port, i.e., there is no knowledge of exact photon number.
For low-intensity output, it is a considerable probability that each photon is assigned to different APD units.
Therefore, the total photon number in each measurement is the sum of all APD trigger counts.
As can be seen from the Fig. \ref{f_detector}(a), to each trigger introduced by single APD there corresponds to an analog voltage of 0.02 V.
We calculate the mean photon number that experimental output, and plot the Poissonian distribution of same photon number.
The theoretical and experimental probability distributions are shown in Fig. \ref{f_detector}(b). 
We use the credibility defined as $H = \sum\nolimits_i  {\sqrt {{x_i}{y_i}} } $ to quantify the
similarity between the experimental probability distribution $\left\{ {x_i}\right\}$ and the theoretical one $\left\{ {y_i} \right\}$ with respect to the Fig. \ref{f_detector}(b).
After calculating we have $H=0.9914$, this implies the detector has a superb credibility.


\begin{thebibliography}{40}%
	\makeatletter
	\providecommand \@ifxundefined [1]{%
		\@ifx{#1\undefined}
	}%
	\providecommand \@ifnum [1]{%
		\ifnum #1\expandafter \@firstoftwo
		\else \expandafter \@secondoftwo
		\fi
	}%
	\providecommand \@ifx [1]{%
		\ifx #1\expandafter \@firstoftwo
		\else \expandafter \@secondoftwo
		\fi
	}%
	\providecommand \natexlab [1]{#1}%
	\providecommand \enquote  [1]{``#1''}%
	\providecommand \bibnamefont  [1]{#1}%
	\providecommand \bibfnamefont [1]{#1}%
	\providecommand \citenamefont [1]{#1}%
	\providecommand \href@noop [0]{\@secondoftwo}%
	\providecommand \href [0]{\begingroup \@sanitize@url \@href}%
	\providecommand \@href[1]{\@@startlink{#1}\@@href}%
	\providecommand \@@href[1]{\endgroup#1\@@endlink}%
	\providecommand \@sanitize@url [0]{\catcode `\\12\catcode `\$12\catcode
		`\&12\catcode `\#12\catcode `\^12\catcode `\_12\catcode `\%12\relax}%
	\providecommand \@@startlink[1]{}%
	\providecommand \@@endlink[0]{}%
	\providecommand \url  [0]{\begingroup\@sanitize@url \@url }%
	\providecommand \@url [1]{\endgroup\@href {#1}{\urlprefix }}%
	\providecommand \urlprefix  [0]{URL }%
	\providecommand \Eprint [0]{\href }%
	\providecommand \doibase [0]{http://dx.doi.org/}%
	\providecommand \selectlanguage [0]{\@gobble}%
	\providecommand \bibinfo  [0]{\@secondoftwo}%
	\providecommand \bibfield  [0]{\@secondoftwo}%
	\providecommand \translation [1]{[#1]}%
	\providecommand \BibitemOpen [0]{}%
	\providecommand \bibitemStop [0]{}%
	\providecommand \bibitemNoStop [0]{.\EOS\space}%
	\providecommand \EOS [0]{\spacefactor3000\relax}%
	\providecommand \BibitemShut  [1]{\csname bibitem#1\endcsname}%
	\let\auto@bib@innerbib\@empty
	%</preamble>
	\bibitem [{\citenamefont {Song}\ \emph {et~al.}(2014)\citenamefont {Song},
		\citenamefont {Lu}, \citenamefont {Gruverman},\ and\ \citenamefont
		{Ducharme}}]{song2014polarization}%
	\BibitemOpen
	\bibfield  {author} {\bibinfo {author} {\bibfnamefont {J.}~\bibnamefont
			{Song}}, \bibinfo {author} {\bibfnamefont {H.}~\bibnamefont {Lu}}, \bibinfo
		{author} {\bibfnamefont {A.}~\bibnamefont {Gruverman}}, \ and\ \bibinfo
		{author} {\bibfnamefont {S.}~\bibnamefont {Ducharme}},\ }\href
	{http://aip.scitation.org/doi/abs/10.1063/1.4875960} {\bibfield  {journal}
		{\bibinfo  {journal} {Appl. Phys. Lett.}\ }\textbf {\bibinfo {volume}
			{104}},\ \bibinfo {pages} {192901} (\bibinfo {year} {2014})}\BibitemShut
	{NoStop}%
	\bibitem [{\citenamefont {Israel}\ \emph {et~al.}(2014)\citenamefont {Israel},
		\citenamefont {Rosen},\ and\ \citenamefont
		{Silberberg}}]{israel2014supersensitive}%
	\BibitemOpen
	\bibfield  {author} {\bibinfo {author} {\bibfnamefont {Y.}~\bibnamefont
			{Israel}}, \bibinfo {author} {\bibfnamefont {S.}~\bibnamefont {Rosen}}, \
		and\ \bibinfo {author} {\bibfnamefont {Y.}~\bibnamefont {Silberberg}},\
	}\href {https://journals.aps.org/prl/abstract/10.1103/PhysRevLett.112.103604}
	{\bibfield  {journal} {\bibinfo  {journal} {Phys. Rev. Lett.}\ }\textbf
		{\bibinfo {volume} {112}},\ \bibinfo {pages} {103604} (\bibinfo {year}
		{2014})}\BibitemShut {NoStop}%
	\bibitem [{\citenamefont {Kartazayeva}\ \emph {et~al.}(2005)\citenamefont
		{Kartazayeva}, \citenamefont {Ni},\ and\ \citenamefont
		{Alfano}}]{kartazayeva2005backscattering}%
	\BibitemOpen
	\bibfield  {author} {\bibinfo {author} {\bibfnamefont {S.}~\bibnamefont
			{Kartazayeva}}, \bibinfo {author} {\bibfnamefont {X.}~\bibnamefont {Ni}}, \
		and\ \bibinfo {author} {\bibfnamefont {R.}~\bibnamefont {Alfano}},\ }\href
	{https://www.osapublishing.org/ol/abstract.cfm?uri=OL-30-10-1168} {\bibfield
		{journal} {\bibinfo  {journal} {Opt. Lett.}\ }\textbf {\bibinfo {volume}
			{30}},\ \bibinfo {pages} {1168} (\bibinfo {year} {2005})}\BibitemShut
	{NoStop}%
	\bibitem [{\citenamefont {Allen}\ \emph {et~al.}(1992)\citenamefont {Allen},
		\citenamefont {Beijersbergen}, \citenamefont {Spreeuw},\ and\ \citenamefont
		{Woerdman}}]{allen1992orbital}%
	\BibitemOpen
	\bibfield  {author} {\bibinfo {author} {\bibfnamefont {L.}~\bibnamefont
			{Allen}}, \bibinfo {author} {\bibfnamefont {M.~W.}\ \bibnamefont
			{Beijersbergen}}, \bibinfo {author} {\bibfnamefont {R.}~\bibnamefont
			{Spreeuw}}, \ and\ \bibinfo {author} {\bibfnamefont {J.}~\bibnamefont
			{Woerdman}},\ }\href
	{https://journals.aps.org/pra/abstract/10.1103/PhysRevA.45.8185} {\bibfield
		{journal} {\bibinfo  {journal} {Phys. Rev. A}\ }\textbf {\bibinfo {volume}
			{45}},\ \bibinfo {pages} {8185} (\bibinfo {year} {1992})}\BibitemShut
	{NoStop}%
	\bibitem [{\citenamefont {Simpson}\ \emph {et~al.}(1997)\citenamefont
		{Simpson}, \citenamefont {Dholakia}, \citenamefont {Allen},\ and\
		\citenamefont {Padgett}}]{simpson1997mechanical}%
	\BibitemOpen
	\bibfield  {author} {\bibinfo {author} {\bibfnamefont {N.}~\bibnamefont
			{Simpson}}, \bibinfo {author} {\bibfnamefont {K.}~\bibnamefont {Dholakia}},
		\bibinfo {author} {\bibfnamefont {L.}~\bibnamefont {Allen}}, \ and\ \bibinfo
		{author} {\bibfnamefont {M.}~\bibnamefont {Padgett}},\ }\href
	{https://www.osapublishing.org/ol/abstract.cfm?uri=ol-22-1-52} {\bibfield
		{journal} {\bibinfo  {journal} {Opt. Lett.}\ }\textbf {\bibinfo {volume}
			{22}},\ \bibinfo {pages} {52} (\bibinfo {year} {1997})}\BibitemShut {NoStop}%
	\bibitem [{\citenamefont {Tabosa}\ and\ \citenamefont
		{Petrov}(1999)}]{tabosa1999optical}%
	\BibitemOpen
	\bibfield  {author} {\bibinfo {author} {\bibfnamefont {J.}~\bibnamefont
			{Tabosa}}\ and\ \bibinfo {author} {\bibfnamefont {D.}~\bibnamefont
			{Petrov}},\ }\href
	{https://journals.aps.org/prl/abstract/10.1103/PhysRevLett.83.4967}
	{\bibfield  {journal} {\bibinfo  {journal} {Phys. Rev. Lett.}\ }\textbf
		{\bibinfo {volume} {83}},\ \bibinfo {pages} {4967} (\bibinfo {year}
		{1999})}\BibitemShut {NoStop}%
	\bibitem [{\citenamefont {Padgett}\ and\ \citenamefont
		{Allen}(2002)}]{padgett2002orbital}%
	\BibitemOpen
	\bibfield  {author} {\bibinfo {author} {\bibfnamefont {M.}~\bibnamefont
			{Padgett}}\ and\ \bibinfo {author} {\bibfnamefont {L.}~\bibnamefont
			{Allen}},\ }\href
	{http://iopscience.iop.org/article/10.1088/1464-4266/4/2/362/meta} {\bibfield
		{journal} {\bibinfo  {journal} {J. Opt. B: Quantum Semicl. Opt.}\ }\textbf
		{\bibinfo {volume} {4}},\ \bibinfo {pages} {S17} (\bibinfo {year}
		{2002})}\BibitemShut {NoStop}%
	\bibitem [{\citenamefont {Yan}\ \emph {et~al.}(2014)\citenamefont {Yan},
		\citenamefont {Xie}, \citenamefont {Lavery}, \citenamefont {Huang},
		\citenamefont {Ahmed}, \citenamefont {Bao}, \citenamefont {Ren},
		\citenamefont {Cao}, \citenamefont {Li}, \citenamefont {Zhao} \emph
		{et~al.}}]{yan2014high}%
	\BibitemOpen
	\bibfield  {author} {\bibinfo {author} {\bibfnamefont {Y.}~\bibnamefont
			{Yan}}, \bibinfo {author} {\bibfnamefont {G.}~\bibnamefont {Xie}}, \bibinfo
		{author} {\bibfnamefont {M.~P.}\ \bibnamefont {Lavery}}, \bibinfo {author}
		{\bibfnamefont {H.}~\bibnamefont {Huang}}, \bibinfo {author} {\bibfnamefont
			{N.}~\bibnamefont {Ahmed}}, \bibinfo {author} {\bibfnamefont
			{C.}~\bibnamefont {Bao}}, \bibinfo {author} {\bibfnamefont {Y.}~\bibnamefont
			{Ren}}, \bibinfo {author} {\bibfnamefont {Y.}~\bibnamefont {Cao}}, \bibinfo
		{author} {\bibfnamefont {L.}~\bibnamefont {Li}}, \bibinfo {author}
		{\bibfnamefont {Z.}~\bibnamefont {Zhao}},  \emph {et~al.},\ }\href
	{https://www.ncbi.nlm.nih.gov/pmc/articles/PMC4175588/} {\bibfield  {journal}
		{\bibinfo  {journal} {Nat. Commun.}\ }\textbf {\bibinfo {volume} {5}}
		(\bibinfo {year} {2014})}\BibitemShut {NoStop}%
	\bibitem [{\citenamefont {Nicolas}\ \emph {et~al.}(2014)\citenamefont
		{Nicolas}, \citenamefont {Veissier}, \citenamefont {Giner}, \citenamefont
		{Giacobino}, \citenamefont {Maxein},\ and\ \citenamefont
		{Laurat}}]{nicolas2014quantum}%
	\BibitemOpen
	\bibfield  {author} {\bibinfo {author} {\bibfnamefont {A.}~\bibnamefont
			{Nicolas}}, \bibinfo {author} {\bibfnamefont {L.}~\bibnamefont {Veissier}},
		\bibinfo {author} {\bibfnamefont {L.}~\bibnamefont {Giner}}, \bibinfo
		{author} {\bibfnamefont {E.}~\bibnamefont {Giacobino}}, \bibinfo {author}
		{\bibfnamefont {D.}~\bibnamefont {Maxein}}, \ and\ \bibinfo {author}
		{\bibfnamefont {J.}~\bibnamefont {Laurat}},\ }\href
	{https://www.nature.com/nphoton/journal/v8/n3/full/nphoton.2013.355.html}
	{\bibfield  {journal} {\bibinfo  {journal} {Nat. Photonics}\ }\textbf
		{\bibinfo {volume} {8}},\ \bibinfo {pages} {234} (\bibinfo {year}
		{2014})}\BibitemShut {NoStop}%
	\bibitem [{\citenamefont {Puentes}\ \emph {et~al.}(2012)\citenamefont
		{Puentes}, \citenamefont {Hermosa},\ and\ \citenamefont
		{Torres}}]{puentes2012weak}%
	\BibitemOpen
	\bibfield  {author} {\bibinfo {author} {\bibfnamefont {G.}~\bibnamefont
			{Puentes}}, \bibinfo {author} {\bibfnamefont {N.}~\bibnamefont {Hermosa}}, \
		and\ \bibinfo {author} {\bibfnamefont {J.}~\bibnamefont {Torres}},\ }\href
	{https://journals.aps.org/prl/abstract/10.1103/PhysRevLett.109.040401}
	{\bibfield  {journal} {\bibinfo  {journal} {Phys. Rev. Lett.}\ }\textbf
		{\bibinfo {volume} {109}},\ \bibinfo {pages} {040401} (\bibinfo {year}
		{2012})}\BibitemShut {NoStop}%
	\bibitem [{\citenamefont {D'ambrosio}\ \emph {et~al.}(2013)\citenamefont
		{D'ambrosio}, \citenamefont {Spagnolo}, \citenamefont {Del~Re}, \citenamefont
		{Slussarenko}, \citenamefont {Li}, \citenamefont {Kwek}, \citenamefont
		{Marrucci}, \citenamefont {Walborn}, \citenamefont {Aolita},\ and\
		\citenamefont {Sciarrino}}]{d2013photonic}%
	\BibitemOpen
	\bibfield  {author} {\bibinfo {author} {\bibfnamefont {V.}~\bibnamefont
			{D'ambrosio}}, \bibinfo {author} {\bibfnamefont {N.}~\bibnamefont
			{Spagnolo}}, \bibinfo {author} {\bibfnamefont {L.}~\bibnamefont {Del~Re}},
		\bibinfo {author} {\bibfnamefont {S.}~\bibnamefont {Slussarenko}}, \bibinfo
		{author} {\bibfnamefont {Y.}~\bibnamefont {Li}}, \bibinfo {author}
		{\bibfnamefont {L.~C.}\ \bibnamefont {Kwek}}, \bibinfo {author}
		{\bibfnamefont {L.}~\bibnamefont {Marrucci}}, \bibinfo {author}
		{\bibfnamefont {S.~P.}\ \bibnamefont {Walborn}}, \bibinfo {author}
		{\bibfnamefont {L.}~\bibnamefont {Aolita}}, \ and\ \bibinfo {author}
		{\bibfnamefont {F.}~\bibnamefont {Sciarrino}},\ }\href
	{https://www.ncbi.nlm.nih.gov/pmc/articles/PMC3791460/} {\bibfield  {journal}
		{\bibinfo  {journal} {Nat. Commun.}\ }\textbf {\bibinfo {volume} {4}}
		(\bibinfo {year} {2013})}\BibitemShut {NoStop}%
	\bibitem [{\citenamefont {Taylor}\ and\ \citenamefont
		{Bowen}(2016)}]{taylor2016quantum}%
	\BibitemOpen
	\bibfield  {author} {\bibinfo {author} {\bibfnamefont {M.~A.}\ \bibnamefont
			{Taylor}}\ and\ \bibinfo {author} {\bibfnamefont {W.~P.}\ \bibnamefont
			{Bowen}},\ }\href
	{http://www.sciencedirect.com/science/article/pii/S0370157315005001}
	{\bibfield  {journal} {\bibinfo  {journal} {Phys. Rep.}\ }\textbf {\bibinfo
			{volume} {615}},\ \bibinfo {pages} {1} (\bibinfo {year} {2016})}\BibitemShut
	{NoStop}%
	\bibitem [{\citenamefont {Giovannetti}\ \emph {et~al.}(2011)\citenamefont
		{Giovannetti}, \citenamefont {Lloyd},\ and\ \citenamefont
		{Maccone}}]{giovannetti2011advances}%
	\BibitemOpen
	\bibfield  {author} {\bibinfo {author} {\bibfnamefont {V.}~\bibnamefont
			{Giovannetti}}, \bibinfo {author} {\bibfnamefont {S.}~\bibnamefont {Lloyd}},
		\ and\ \bibinfo {author} {\bibfnamefont {L.}~\bibnamefont {Maccone}},\ }\href
	{http://www.nature.com/nphoton/journal/v5/n4/full/nphoton.2011.35.html?foxtrotcallback=true}
	{\bibfield  {journal} {\bibinfo  {journal} {Nat. Photonics}\ }\textbf
		{\bibinfo {volume} {5}},\ \bibinfo {pages} {222} (\bibinfo {year}
		{2011})}\BibitemShut {NoStop}%
	\bibitem [{\citenamefont {Zhou}\ \emph {et~al.}(2015)\citenamefont {Zhou},
		\citenamefont {Cable}, \citenamefont {Whittaker}, \citenamefont {Shadbolt},
		\citenamefont {O��Brien},\ and\ \citenamefont
		{Matthews}}]{zhou2015quantum}%
	\BibitemOpen
	\bibfield  {author} {\bibinfo {author} {\bibfnamefont {X.-Q.}\ \bibnamefont
			{Zhou}}, \bibinfo {author} {\bibfnamefont {H.}~\bibnamefont {Cable}},
		\bibinfo {author} {\bibfnamefont {R.}~\bibnamefont {Whittaker}}, \bibinfo
		{author} {\bibfnamefont {P.}~\bibnamefont {Shadbolt}}, \bibinfo {author}
		{\bibfnamefont {J.~L.}\ \bibnamefont {O��Brien}}, \ and\ \bibinfo
		{author} {\bibfnamefont {J.~C.}\ \bibnamefont {Matthews}},\ }\href
	{https://www.osapublishing.org/optica/abstract.cfm?uri=optica-2-6-510}
	{\bibfield  {journal} {\bibinfo  {journal} {Optica}\ }\textbf {\bibinfo
			{volume} {2}},\ \bibinfo {pages} {510} (\bibinfo {year} {2015})}\BibitemShut
	{NoStop}%
	\bibitem [{\citenamefont {Maga{\~n}a-Loaiza}\ \emph {et~al.}(2014)\citenamefont
		{Maga{\~n}a-Loaiza}, \citenamefont {Mirhosseini}, \citenamefont {Rodenburg},\
		and\ \citenamefont {Boyd}}]{magana2014amplification}%
	\BibitemOpen
	\bibfield  {author} {\bibinfo {author} {\bibfnamefont {O.~S.}\ \bibnamefont
			{Maga{\~n}a-Loaiza}}, \bibinfo {author} {\bibfnamefont {M.}~\bibnamefont
			{Mirhosseini}}, \bibinfo {author} {\bibfnamefont {B.}~\bibnamefont
			{Rodenburg}}, \ and\ \bibinfo {author} {\bibfnamefont {R.~W.}\ \bibnamefont
			{Boyd}},\ }\href
	{https://journals.aps.org/prl/abstract/10.1103/PhysRevLett.112.200401}
	{\bibfield  {journal} {\bibinfo  {journal} {Phys. Rev. Lett.}\ }\textbf
		{\bibinfo {volume} {112}},\ \bibinfo {pages} {200401} (\bibinfo {year}
		{2014})}\BibitemShut {NoStop}%
	\bibitem [{\citenamefont {Thekkadath}\ \emph {et~al.}(2016)\citenamefont
		{Thekkadath}, \citenamefont {Giner}, \citenamefont {Chalich}, \citenamefont
		{Horton}, \citenamefont {Banker},\ and\ \citenamefont
		{Lundeen}}]{thekkadath2016direct}%
	\BibitemOpen
	\bibfield  {author} {\bibinfo {author} {\bibfnamefont {G.}~\bibnamefont
			{Thekkadath}}, \bibinfo {author} {\bibfnamefont {L.}~\bibnamefont {Giner}},
		\bibinfo {author} {\bibfnamefont {Y.}~\bibnamefont {Chalich}}, \bibinfo
		{author} {\bibfnamefont {M.}~\bibnamefont {Horton}}, \bibinfo {author}
		{\bibfnamefont {J.}~\bibnamefont {Banker}}, \ and\ \bibinfo {author}
		{\bibfnamefont {J.}~\bibnamefont {Lundeen}},\ }\href
	{https://journals.aps.org/prl/abstract/10.1103/PhysRevLett.117.120401}
	{\bibfield  {journal} {\bibinfo  {journal} {Phys. Rev. Lett.}\ }\textbf
		{\bibinfo {volume} {117}},\ \bibinfo {pages} {120401} (\bibinfo {year}
		{2016})}\BibitemShut {NoStop}%
	\bibitem [{\citenamefont {Zhang}\ \emph {et~al.}(2015)\citenamefont {Zhang},
		\citenamefont {Datta},\ and\ \citenamefont {Walmsley}}]{zhang2015precision}%
	\BibitemOpen
	\bibfield  {author} {\bibinfo {author} {\bibfnamefont {L.}~\bibnamefont
			{Zhang}}, \bibinfo {author} {\bibfnamefont {A.}~\bibnamefont {Datta}}, \ and\
		\bibinfo {author} {\bibfnamefont {I.~A.}\ \bibnamefont {Walmsley}},\ }\href
	{https://journals.aps.org/prl/abstract/10.1103/PhysRevLett.114.210801}
	{\bibfield  {journal} {\bibinfo  {journal} {Phys. Rev. Lett.}\ }\textbf
		{\bibinfo {volume} {114}},\ \bibinfo {pages} {210801} (\bibinfo {year}
		{2015})}\BibitemShut {NoStop}%
	\bibitem [{\citenamefont {Fickler}\ \emph {et~al.}(2012)\citenamefont
		{Fickler}, \citenamefont {Lapkiewicz}, \citenamefont {Plick}, \citenamefont
		{Krenn}, \citenamefont {Schaeff}, \citenamefont {Ramelow},\ and\
		\citenamefont {Zeilinger}}]{fickler2012quantum}%
	\BibitemOpen
	\bibfield  {author} {\bibinfo {author} {\bibfnamefont {R.}~\bibnamefont
			{Fickler}}, \bibinfo {author} {\bibfnamefont {R.}~\bibnamefont {Lapkiewicz}},
		\bibinfo {author} {\bibfnamefont {W.~N.}\ \bibnamefont {Plick}}, \bibinfo
		{author} {\bibfnamefont {M.}~\bibnamefont {Krenn}}, \bibinfo {author}
		{\bibfnamefont {C.}~\bibnamefont {Schaeff}}, \bibinfo {author} {\bibfnamefont
			{S.}~\bibnamefont {Ramelow}}, \ and\ \bibinfo {author} {\bibfnamefont
			{A.}~\bibnamefont {Zeilinger}},\ }\href
	{http://science.sciencemag.org/content/338/6107/640} {\bibfield  {journal}
		{\bibinfo  {journal} {Science}\ }\textbf {\bibinfo {volume} {338}},\ \bibinfo
		{pages} {640} (\bibinfo {year} {2012})}\BibitemShut {NoStop}%
	\bibitem [{\citenamefont {Zhou}\ \emph {et~al.}(2016)\citenamefont {Zhou},
		\citenamefont {Liu}, \citenamefont {Li}, \citenamefont {Ding}, \citenamefont
		{Zhang}, \citenamefont {Shi}, \citenamefont {Dong}, \citenamefont {Shi},\
		and\ \citenamefont {Guo}}]{zhou2016orbital}%
	\BibitemOpen
	\bibfield  {author} {\bibinfo {author} {\bibfnamefont {Z.-Y.}\ \bibnamefont
			{Zhou}}, \bibinfo {author} {\bibfnamefont {S.-L.}\ \bibnamefont {Liu}},
		\bibinfo {author} {\bibfnamefont {Y.}~\bibnamefont {Li}}, \bibinfo {author}
		{\bibfnamefont {D.-S.}\ \bibnamefont {Ding}}, \bibinfo {author}
		{\bibfnamefont {W.}~\bibnamefont {Zhang}}, \bibinfo {author} {\bibfnamefont
			{S.}~\bibnamefont {Shi}}, \bibinfo {author} {\bibfnamefont {M.-X.}\
			\bibnamefont {Dong}}, \bibinfo {author} {\bibfnamefont {B.-S.}\ \bibnamefont
			{Shi}}, \ and\ \bibinfo {author} {\bibfnamefont {G.-C.}\ \bibnamefont
			{Guo}},\ }\href
	{https://journals.aps.org/prl/abstract/10.1103/PhysRevLett.117.103601}
	{\bibfield  {journal} {\bibinfo  {journal} {Phys. Rev. Lett.}\ }\textbf
		{\bibinfo {volume} {117}},\ \bibinfo {pages} {103601} (\bibinfo {year}
		{2016})}\BibitemShut {NoStop}%
	\bibitem [{\citenamefont {Leach}\ \emph {et~al.}(2002)\citenamefont {Leach},
		\citenamefont {Padgett}, \citenamefont {Barnett}, \citenamefont
		{Franke-Arnold},\ and\ \citenamefont {Courtial}}]{leach2002measuring}%
	\BibitemOpen
	\bibfield  {author} {\bibinfo {author} {\bibfnamefont {J.}~\bibnamefont
			{Leach}}, \bibinfo {author} {\bibfnamefont {M.~J.}\ \bibnamefont {Padgett}},
		\bibinfo {author} {\bibfnamefont {S.~M.}\ \bibnamefont {Barnett}}, \bibinfo
		{author} {\bibfnamefont {S.}~\bibnamefont {Franke-Arnold}}, \ and\ \bibinfo
		{author} {\bibfnamefont {J.}~\bibnamefont {Courtial}},\ }\href
	{https://journals.aps.org/prl/abstract/10.1103/PhysRevLett.88.257901}
	{\bibfield  {journal} {\bibinfo  {journal} {Phys. Rev. Lett.}\ }\textbf
		{\bibinfo {volume} {88}},\ \bibinfo {pages} {257901} (\bibinfo {year}
		{2002})}\BibitemShut {NoStop}%
	\bibitem [{\citenamefont {Bollinger}\ \emph {et~al.}(1996)\citenamefont
		{Bollinger}, \citenamefont {Itano}, \citenamefont {Wineland},\ and\
		\citenamefont {Heinzen}}]{bollinger1996optimal}%
	\BibitemOpen
	\bibfield  {author} {\bibinfo {author} {\bibfnamefont {J.~J.}\ \bibnamefont
			{Bollinger}}, \bibinfo {author} {\bibfnamefont {W.~M.}\ \bibnamefont
			{Itano}}, \bibinfo {author} {\bibfnamefont {D.~J.}\ \bibnamefont {Wineland}},
		\ and\ \bibinfo {author} {\bibfnamefont {D.}~\bibnamefont {Heinzen}},\ }\href
	{https://journals.aps.org/pra/abstract/10.1103/PhysRevA.54.R4649} {\bibfield
		{journal} {\bibinfo  {journal} {Phys. Rev. A}\ }\textbf {\bibinfo {volume}
			{54}},\ \bibinfo {pages} {R4649} (\bibinfo {year} {1996})}\BibitemShut
	{NoStop}%
	\bibitem [{\citenamefont {Gerry}(2000)}]{gerry2000heisenberg}%
	\BibitemOpen
	\bibfield  {author} {\bibinfo {author} {\bibfnamefont {C.~C.}\ \bibnamefont
			{Gerry}},\ }\href
	{https://journals.aps.org/pra/abstract/10.1103/PhysRevA.61.043811} {\bibfield
		{journal} {\bibinfo  {journal} {Phys. Rev. A}\ }\textbf {\bibinfo {volume}
			{61}},\ \bibinfo {pages} {043811} (\bibinfo {year} {2000})}\BibitemShut
	{NoStop}%
	\bibitem [{\citenamefont {Gerry}\ \emph {et~al.}(2005)\citenamefont {Gerry},
		\citenamefont {Benmoussa},\ and\ \citenamefont {Campos}}]{gerry2005quantum}%
	\BibitemOpen
	\bibfield  {author} {\bibinfo {author} {\bibfnamefont {C.~C.}\ \bibnamefont
			{Gerry}}, \bibinfo {author} {\bibfnamefont {A.}~\bibnamefont {Benmoussa}}, \
		and\ \bibinfo {author} {\bibfnamefont {R.}~\bibnamefont {Campos}},\ }\href
	{https://journals.aps.org/pra/abstract/10.1103/PhysRevA.72.053818} {\bibfield
		{journal} {\bibinfo  {journal} {Phys. Rev. A}\ }\textbf {\bibinfo {volume}
			{72}},\ \bibinfo {pages} {053818} (\bibinfo {year} {2005})}\BibitemShut
	{NoStop}%
	\bibitem [{\citenamefont {Achilles}\ \emph {et~al.}(2004)\citenamefont
		{Achilles}, \citenamefont {Silberhorn}, \citenamefont {Sliwa}, \citenamefont
		{Banaszek}, \citenamefont {Walmsley}, \citenamefont {Fitch}, \citenamefont
		{Jacobs}, \citenamefont {Pittman},\ and\ \citenamefont
		{Franson}}]{achilles2004photon}%
	\BibitemOpen
	\bibfield  {author} {\bibinfo {author} {\bibfnamefont {D.}~\bibnamefont
			{Achilles}}, \bibinfo {author} {\bibfnamefont {C.}~\bibnamefont
			{Silberhorn}}, \bibinfo {author} {\bibfnamefont {C.}~\bibnamefont {Sliwa}},
		\bibinfo {author} {\bibfnamefont {K.}~\bibnamefont {Banaszek}}, \bibinfo
		{author} {\bibfnamefont {I.~A.}\ \bibnamefont {Walmsley}}, \bibinfo {author}
		{\bibfnamefont {M.~J.}\ \bibnamefont {Fitch}}, \bibinfo {author}
		{\bibfnamefont {B.~C.}\ \bibnamefont {Jacobs}}, \bibinfo {author}
		{\bibfnamefont {T.~B.}\ \bibnamefont {Pittman}}, \ and\ \bibinfo {author}
		{\bibfnamefont {J.~D.}\ \bibnamefont {Franson}},\ }\href
	{http://www.tandfonline.com/doi/abs/10.1080/09500340408235288} {\bibfield
		{journal} {\bibinfo  {journal} {J. Mod. Opt.}\ }\textbf {\bibinfo {volume}
			{51}},\ \bibinfo {pages} {1499} (\bibinfo {year} {2004})}\BibitemShut
	{NoStop}%
	\bibitem [{\citenamefont {Cohen}\ \emph {et~al.}(2014)\citenamefont {Cohen},
		\citenamefont {Istrati}, \citenamefont {Dovrat},\ and\ \citenamefont
		{Eisenberg}}]{cohen2014super}%
	\BibitemOpen
	\bibfield  {author} {\bibinfo {author} {\bibfnamefont {L.}~\bibnamefont
			{Cohen}}, \bibinfo {author} {\bibfnamefont {D.}~\bibnamefont {Istrati}},
		\bibinfo {author} {\bibfnamefont {L.}~\bibnamefont {Dovrat}}, \ and\ \bibinfo
		{author} {\bibfnamefont {H.}~\bibnamefont {Eisenberg}},\ }\href
	{https://www.osapublishing.org/oe/abstract.cfm?uri=oe-22-10-11945} {\bibfield
		{journal} {\bibinfo  {journal} {Opt. Express}\ }\textbf {\bibinfo {volume}
			{22}},\ \bibinfo {pages} {11945} (\bibinfo {year} {2014})}\BibitemShut
	{NoStop}%
	\bibitem [{\citenamefont {Liu}\ \emph {et~al.}(2017)\citenamefont {Liu},
		\citenamefont {Wang}, \citenamefont {Yang}, \citenamefont {Jin},\ and\
		\citenamefont {Sun}}]{liu2017fisher}%
	\BibitemOpen
	\bibfield  {author} {\bibinfo {author} {\bibfnamefont {P.}~\bibnamefont
			{Liu}}, \bibinfo {author} {\bibfnamefont {P.}~\bibnamefont {Wang}}, \bibinfo
		{author} {\bibfnamefont {W.}~\bibnamefont {Yang}}, \bibinfo {author}
		{\bibfnamefont {G.}~\bibnamefont {Jin}}, \ and\ \bibinfo {author}
		{\bibfnamefont {C.}~\bibnamefont {Sun}},\ }\href
	{https://journals.aps.org/pra/abstract/10.1103/PhysRevA.95.023824} {\bibfield
		{journal} {\bibinfo  {journal} {Phys. Rev. A}\ }\textbf {\bibinfo {volume}
			{95}},\ \bibinfo {pages} {023824} (\bibinfo {year} {2017})}\BibitemShut
	{NoStop}%
	\bibitem [{\citenamefont {Dowling}(2008)}]{N00N}%
	\BibitemOpen
	\bibfield  {author} {\bibinfo {author} {\bibfnamefont {J.~P.}\ \bibnamefont
			{Dowling}},\ }\href {\doibase 10.1080/00107510802091298} {\bibfield
		{journal} {\bibinfo  {journal} {Contemp. Phys.}\ }\textbf {\bibinfo {volume}
			{49}},\ \bibinfo {pages} {125} (\bibinfo {year} {2008})}\BibitemShut
	{NoStop}%
	\bibitem [{\citenamefont {Bahder}(2011)}]{bahder2011phase}%
	\BibitemOpen
	\bibfield  {author} {\bibinfo {author} {\bibfnamefont {T.~B.}\ \bibnamefont
			{Bahder}},\ }\href
	{https://journals.aps.org/pra/abstract/10.1103/PhysRevA.83.053601} {\bibfield
		{journal} {\bibinfo  {journal} {Phys. Rev. A}\ }\textbf {\bibinfo {volume}
			{83}},\ \bibinfo {pages} {053601} (\bibinfo {year} {2011})}\BibitemShut
	{NoStop}%
	\bibitem [{\citenamefont {Zhang}\ \emph {et~al.}(2017)\citenamefont {Zhang},
		\citenamefont {Zhang}, \citenamefont {Cen}, \citenamefont {Yu}, \citenamefont
		{Li}, \citenamefont {Wang},\ and\ \citenamefont {Zhao}}]{zhang17effects}%
	\BibitemOpen
	\bibfield  {author} {\bibinfo {author} {\bibfnamefont {J.}~\bibnamefont
			{Zhang}}, \bibinfo {author} {\bibfnamefont {Z.}~\bibnamefont {Zhang}},
		\bibinfo {author} {\bibfnamefont {L.}~\bibnamefont {Cen}}, \bibinfo {author}
		{\bibfnamefont {M.}~\bibnamefont {Yu}}, \bibinfo {author} {\bibfnamefont
			{S.}~\bibnamefont {Li}}, \bibinfo {author} {\bibfnamefont {F.}~\bibnamefont
			{Wang}}, \ and\ \bibinfo {author} {\bibfnamefont {Y.}~\bibnamefont {Zhao}},\
	}\href {https://www.osapublishing.org/oe/abstract.cfm?uri=oe-25-21-24907}
	{\bibfield  {journal} {\bibinfo  {journal} {Opt. Express}\ }\textbf {\bibinfo
			{volume} {25}},\ \bibinfo {pages} {24907} (\bibinfo {year}
		{2017})}\BibitemShut {NoStop}%
	\bibitem [{\citenamefont {Feng}\ \emph {et~al.}(2014)\citenamefont {Feng},
		\citenamefont {Jin},\ and\ \citenamefont {Yang}}]{feng2014quantum}%
	\BibitemOpen
	\bibfield  {author} {\bibinfo {author} {\bibfnamefont {X.}~\bibnamefont
			{Feng}}, \bibinfo {author} {\bibfnamefont {G.}~\bibnamefont {Jin}}, \ and\
		\bibinfo {author} {\bibfnamefont {W.}~\bibnamefont {Yang}},\ }\href
	{https://journals.aps.org/pra/abstract/10.1103/PhysRevA.90.013807} {\bibfield
		{journal} {\bibinfo  {journal} {Phys. Rev. A}\ }\textbf {\bibinfo {volume}
			{90}},\ \bibinfo {pages} {013807} (\bibinfo {year} {2014})}\BibitemShut
	{NoStop}%
	\bibitem [{\citenamefont {Kacprowicz}\ \emph {et~al.}(2010)\citenamefont
		{Kacprowicz}, \citenamefont {Demkowicz-Dobrza{\'n}ski}, \citenamefont
		{Wasilewski}, \citenamefont {Banaszek},\ and\ \citenamefont
		{Walmsley}}]{kacprowicz2010experimental}%
	\BibitemOpen
	\bibfield  {author} {\bibinfo {author} {\bibfnamefont {M.}~\bibnamefont
			{Kacprowicz}}, \bibinfo {author} {\bibfnamefont {R.}~\bibnamefont
			{Demkowicz-Dobrza{\'n}ski}}, \bibinfo {author} {\bibfnamefont
			{W.}~\bibnamefont {Wasilewski}}, \bibinfo {author} {\bibfnamefont
			{K.}~\bibnamefont {Banaszek}}, \ and\ \bibinfo {author} {\bibfnamefont
			{I.}~\bibnamefont {Walmsley}},\ }\href
	{https://www.nature.com/nphoton/journal/v4/n6/full/nphoton.2010.39.html}
	{\bibfield  {journal} {\bibinfo  {journal} {Nat. Photonics}\ }\textbf
		{\bibinfo {volume} {4}},\ \bibinfo {pages} {357} (\bibinfo {year}
		{2010})}\BibitemShut {NoStop}%
	\bibitem [{\citenamefont {Spagnolo}\ \emph {et~al.}(2012)\citenamefont
		{Spagnolo}, \citenamefont {Vitelli}, \citenamefont {Lucivero}, \citenamefont
		{Giovannetti}, \citenamefont {Maccone},\ and\ \citenamefont
		{Sciarrino}}]{spagnolo2012phase}%
	\BibitemOpen
	\bibfield  {author} {\bibinfo {author} {\bibfnamefont {N.}~\bibnamefont
			{Spagnolo}}, \bibinfo {author} {\bibfnamefont {C.}~\bibnamefont {Vitelli}},
		\bibinfo {author} {\bibfnamefont {V.~G.}\ \bibnamefont {Lucivero}}, \bibinfo
		{author} {\bibfnamefont {V.}~\bibnamefont {Giovannetti}}, \bibinfo {author}
		{\bibfnamefont {L.}~\bibnamefont {Maccone}}, \ and\ \bibinfo {author}
		{\bibfnamefont {F.}~\bibnamefont {Sciarrino}},\ }\href
	{https://journals.aps.org/prl/abstract/10.1103/PhysRevLett.108.233602}
	{\bibfield  {journal} {\bibinfo  {journal} {Phys. Rev. Lett.}\ }\textbf
		{\bibinfo {volume} {108}},\ \bibinfo {pages} {233602} (\bibinfo {year}
		{2012})}\BibitemShut {NoStop}%
	\bibitem [{\citenamefont {Datta}\ \emph {et~al.}(2011)\citenamefont {Datta},
		\citenamefont {Zhang}, \citenamefont {Thomas-Peter}, \citenamefont {Dorner},
		\citenamefont {Smith},\ and\ \citenamefont {Walmsley}}]{PhysRevA.83.063836}%
	\BibitemOpen
	\bibfield  {author} {\bibinfo {author} {\bibfnamefont {A.}~\bibnamefont
			{Datta}}, \bibinfo {author} {\bibfnamefont {L.}~\bibnamefont {Zhang}},
		\bibinfo {author} {\bibfnamefont {N.}~\bibnamefont {Thomas-Peter}}, \bibinfo
		{author} {\bibfnamefont {U.}~\bibnamefont {Dorner}}, \bibinfo {author}
		{\bibfnamefont {B.~J.}\ \bibnamefont {Smith}}, \ and\ \bibinfo {author}
		{\bibfnamefont {I.~A.}\ \bibnamefont {Walmsley}},\ }\href {\doibase
		10.1103/PhysRevA.83.063836} {\bibfield  {journal} {\bibinfo  {journal} {Phys.
				Rev. A}\ }\textbf {\bibinfo {volume} {83}},\ \bibinfo {pages} {063836}
		(\bibinfo {year} {2011})}\BibitemShut {NoStop}%
	\bibitem [{\citenamefont {Pezz\'e}\ \emph {et~al.}(2007)\citenamefont
		{Pezz\'e}, \citenamefont {Smerzi}, \citenamefont {Khoury}, \citenamefont
		{Hodelin},\ and\ \citenamefont {Bouwmeester}}]{PhysRevLett.99.223602}%
	\BibitemOpen
	\bibfield  {author} {\bibinfo {author} {\bibfnamefont {L.}~\bibnamefont
			{Pezz\'e}}, \bibinfo {author} {\bibfnamefont {A.}~\bibnamefont {Smerzi}},
		\bibinfo {author} {\bibfnamefont {G.}~\bibnamefont {Khoury}}, \bibinfo
		{author} {\bibfnamefont {J.~F.}\ \bibnamefont {Hodelin}}, \ and\ \bibinfo
		{author} {\bibfnamefont {D.}~\bibnamefont {Bouwmeester}},\ }\href {\doibase
		10.1103/PhysRevLett.99.223602} {\bibfield  {journal} {\bibinfo  {journal}
			{Phys. Rev. Lett.}\ }\textbf {\bibinfo {volume} {99}},\ \bibinfo {pages}
		{223602} (\bibinfo {year} {2007})}\BibitemShut {NoStop}%
	\bibitem [{\citenamefont {Huang}\ \emph {et~al.}(2017)\citenamefont {Huang},
		\citenamefont {Motes}, \citenamefont {Anisimov}, \citenamefont {Dowling},\
		and\ \citenamefont {Berry}}]{huang2017adaptive}%
	\BibitemOpen
	\bibfield  {author} {\bibinfo {author} {\bibfnamefont {Z.}~\bibnamefont
			{Huang}}, \bibinfo {author} {\bibfnamefont {K.~R.}\ \bibnamefont {Motes}},
		\bibinfo {author} {\bibfnamefont {P.~M.}\ \bibnamefont {Anisimov}}, \bibinfo
		{author} {\bibfnamefont {J.~P.}\ \bibnamefont {Dowling}}, \ and\ \bibinfo
		{author} {\bibfnamefont {D.~W.}\ \bibnamefont {Berry}},\ }\href
	{https://journals.aps.org/pra/abstract/10.1103/PhysRevA.95.053837} {\bibfield
		{journal} {\bibinfo  {journal} {Phys. Rev. A}\ }\textbf {\bibinfo {volume}
			{95}},\ \bibinfo {pages} {053837} (\bibinfo {year} {2017})}\BibitemShut
	{NoStop}%
	\bibitem [{\citenamefont {Tan}\ \emph {et~al.}(2014)\citenamefont {Tan},
		\citenamefont {Liao}, \citenamefont {Wang},\ and\ \citenamefont
		{Nori}}]{tan2014enhanced}%
	\BibitemOpen
	\bibfield  {author} {\bibinfo {author} {\bibfnamefont {Q.-S.}\ \bibnamefont
			{Tan}}, \bibinfo {author} {\bibfnamefont {J.-Q.}\ \bibnamefont {Liao}},
		\bibinfo {author} {\bibfnamefont {X.}~\bibnamefont {Wang}}, \ and\ \bibinfo
		{author} {\bibfnamefont {F.}~\bibnamefont {Nori}},\ }\href
	{https://journals.aps.org/pra/abstract/10.1103/PhysRevA.89.053822} {\bibfield
		{journal} {\bibinfo  {journal} {Phys. Rev. A}\ }\textbf {\bibinfo {volume}
			{89}},\ \bibinfo {pages} {053822} (\bibinfo {year} {2014})}\BibitemShut
	{NoStop}%
	\bibitem [{\citenamefont {Gibilisco}\ \emph {et~al.}(2007)\citenamefont
		{Gibilisco}, \citenamefont {Imparato},\ and\ \citenamefont
		{Isola}}]{gibilisco2007uncertainty}%
	\BibitemOpen
	\bibfield  {author} {\bibinfo {author} {\bibfnamefont {P.}~\bibnamefont
			{Gibilisco}}, \bibinfo {author} {\bibfnamefont {D.}~\bibnamefont {Imparato}},
		\ and\ \bibinfo {author} {\bibfnamefont {T.}~\bibnamefont {Isola}},\ }\href
	{http://aip.scitation.org/doi/abs/10.1063/1.2748210} {\bibfield  {journal}
		{\bibinfo  {journal} {J. Math. Phys.}\ }\textbf {\bibinfo {volume} {48}},\
		\bibinfo {pages} {072109} (\bibinfo {year} {2007})}\BibitemShut {NoStop}%
	\bibitem [{\citenamefont {Knysh}\ \emph {et~al.}(2011)\citenamefont {Knysh},
		\citenamefont {Smelyanskiy},\ and\ \citenamefont
		{Durkin}}]{knysh2011scaling}%
	\BibitemOpen
	\bibfield  {author} {\bibinfo {author} {\bibfnamefont {S.}~\bibnamefont
			{Knysh}}, \bibinfo {author} {\bibfnamefont {V.~N.}\ \bibnamefont
			{Smelyanskiy}}, \ and\ \bibinfo {author} {\bibfnamefont {G.~A.}\ \bibnamefont
			{Durkin}},\ }\href
	{https://journals.aps.org/pra/abstract/10.1103/PhysRevA.83.021804} {\bibfield
		{journal} {\bibinfo  {journal} {Phys. Rev. A}\ }\textbf {\bibinfo {volume}
			{83}},\ \bibinfo {pages} {021804} (\bibinfo {year} {2011})}\BibitemShut
	{NoStop}%
	\bibitem [{\citenamefont {Liu}\ \emph {et~al.}(2013)\citenamefont {Liu},
		\citenamefont {Jing},\ and\ \citenamefont {Wang}}]{liu2013phase}%
	\BibitemOpen
	\bibfield  {author} {\bibinfo {author} {\bibfnamefont {J.}~\bibnamefont
			{Liu}}, \bibinfo {author} {\bibfnamefont {X.}~\bibnamefont {Jing}}, \ and\
		\bibinfo {author} {\bibfnamefont {X.}~\bibnamefont {Wang}},\ }\href
	{https://journals.aps.org/pra/abstract/10.1103/PhysRevA.88.042316} {\bibfield
		{journal} {\bibinfo  {journal} {Phys. Rev. A}\ }\textbf {\bibinfo {volume}
			{88}},\ \bibinfo {pages} {042316} (\bibinfo {year} {2013})}\BibitemShut
	{NoStop}%
	\bibitem [{\citenamefont {Boto}\ \emph {et~al.}(2000)\citenamefont {Boto},
		\citenamefont {Kok}, \citenamefont {Abrams}, \citenamefont {Braunstein},
		\citenamefont {Williams},\ and\ \citenamefont
		{Dowling}}]{PhysRevLett.85.2733}%
	\BibitemOpen
	\bibfield  {author} {\bibinfo {author} {\bibfnamefont {A.~N.}\ \bibnamefont
			{Boto}}, \bibinfo {author} {\bibfnamefont {P.}~\bibnamefont {Kok}}, \bibinfo
		{author} {\bibfnamefont {D.~S.}\ \bibnamefont {Abrams}}, \bibinfo {author}
		{\bibfnamefont {S.~L.}\ \bibnamefont {Braunstein}}, \bibinfo {author}
		{\bibfnamefont {C.~P.}\ \bibnamefont {Williams}}, \ and\ \bibinfo {author}
		{\bibfnamefont {J.~P.}\ \bibnamefont {Dowling}},\ }\href {\doibase
		10.1103/PhysRevLett.85.2733} {\bibfield  {journal} {\bibinfo  {journal}
			{Phys. Rev. Lett.}\ }\textbf {\bibinfo {volume} {85}},\ \bibinfo {pages}
		{2733} (\bibinfo {year} {2000})}\BibitemShut {NoStop}%
\end{thebibliography}
\end{document}